\documentclass[aps,reprint,nofootinbib,superscriptaddress]{revtex4-2}

\providecommand{\sorthelp}[1]{}
 
\usepackage{graphicx}
\usepackage{amssymb}
\usepackage{amsmath}
\usepackage{epstopdf}
\usepackage{multirow}
\usepackage[colorlinks=true,linkcolor=blue,citecolor=blue]{hyperref}%
\usepackage{orcidlink} 
\usepackage{graphicx}
\usepackage{bm}
\usepackage{xcolor}
\usepackage{siunitx}

\usepackage[caption=false]{subfig}
\usepackage[capitalise]{cleveref}
\usepackage{mhchem}

\newcommand{\be}{\begin{eqnarray}}

\newcommand{\ee}{\end{eqnarray}}

\usepackage[normalem]{ulem}

\usepackage[export]{adjustbox}% http://ctan.org/pkg/adjustbox

\providecommand{\sorthelp}[1]{}
\begin{document}

\preprint{APS/123-Q ED}

\title{The Atacama Cosmology Telescope: A measurement of galaxy cluster temperatures through relativistic corrections to the thermal Sunyaev-Zeldovich effect} 

\newcommand{\cca}{CCA}

\author{William~R.~Coulton\orcidlink{0000-0002-1297-3673}}
\affiliation{Kavli Institute for Cosmology Cambridge, Madingley Road, Cambridge CB3 0HA, UK}
\affiliation{DAMTP, Centre for Mathematical Sciences, University of Cambridge, Wilberforce Road, Cambridge CB3 OWA, UK}
\affiliation{Center for Computational Astrophysics, Flatiron Institute, 162 5th Avenue, New York, NY 10010 USA}
\author{Adriaan~J.~Duivenvoorden}\affiliation{Max-Planck-Institut f\"{u}r Astrophysik, Karl-Schwarzschild-Str.\ 1, 85748 Garching, Germany}\affiliation{Center for Computational Astrophysics, Flatiron Institute, 162 5th Avenue, New York, NY 10010 USA}\affiliation{Joseph Henry Laboratories of Physics, Jadwin Hall,
Princeton University, Princeton, NJ, USA 08544}
\author{Zachary~Atkins\orcidlink{0000-0002-2287-1603}}
\affiliation{Joseph Henry Laboratories of Physics, Jadwin Hall, Princeton University, Princeton, NJ, USA 08544}
\author{Nicholas~Battaglia}\affiliation{Department of Astronomy, Cornell University, Ithaca, NY 14853, USA}
\author{Elia~Stefano~Battistelli}\affiliation{Sapienza University of Rome, Physics Department, Piazzale Aldo Moro 5, 00185 Rome, Italy}
\author{J~Richard~Bond}\affiliation{Canadian Institute for Theoretical Astrophysics, University of
Toronto, Toronto, ON, Canada M5S 3H8}
\author{Hongbo Cai \orcidlink{
[0000-0003-3851-7518}}
\affiliation{Department of Physics and Astronomy, University of
Pittsburgh, Pittsburgh, PA,
USA 15260}
\author{Erminia~Calabrese}\affiliation{School of Physics and Astronomy, Cardiff University, The Parade, 
Cardiff, Wales, UK CF24 3AA}
\author{Steve~K.~Choi\orcidlink{0000-0002-9113-7058}}
\affiliation{Department of Physics and Astronomy, University of California, Riverside, CA 92521, USA}
\author{Kevin T. Crowley}\affiliation{Department of Astronomy and Astrophysics, University of California San Diego, La Jolla, CA 92093, USA}
\author{Mark~J.~Devlin}\affiliation{Department of Physics and Astronomy, University of
Pennsylvania, 209 South 33rd Street, Philadelphia, PA, USA 19104}
\author{Jo~Dunkley}\affiliation{Joseph Henry Laboratories of Physics, Jadwin Hall,
Princeton University, Princeton, NJ, USA 08544}\affiliation{Department of Astrophysical Sciences, Peyton Hall, 
Princeton University, Princeton, NJ USA 08544}
\author{Simone Ferraro\orcidlink{0000-0003-4992-7854}}
\affiliation{Lawrence Berkeley National Laboratory, One Cyclotron Road, Berkeley, CA 94720, USA}
\affiliation{Berkeley Center for Cosmological Physics, Department of Physics,
University of California, Berkeley, CA 94720, USA}
\author{Yilun Guan\orcidlink{
ORCID: 0000-0002-1697-3080}}
\affiliation{Dunlap Institute for Astronomy \& Astrophysics, University of Toronto, 50 St. George St., Toronto ON M5S 3H4, Canada}
\author{Carlos~Herv\'ias-Caimapo}\affiliation{Instituto de Astrof\'isica and Centro de Astro-Ingenier\'ia, Facultad de F\`isica, Pontificia Universidad Cat\'olica de Chile, Av. Vicu\~na Mackenna 4860, 7820436 Macul, Santiago, Chile}
\author{J.~Colin~Hill}\affiliation{Department of Physics, Columbia University, New York, NY, USA}\affiliation{Center for Computational Astrophysics, Flatiron Institute, 162 5th Avenue, New York, NY 10010 USA}
\author{Matt~Hilton}\affiliation{Wits Centre for Astrophysics, School of Physics, University of the Witwatersrand, Private Bag 3, 2050, Johannesburg, South Africa}\affiliation{Astrophysics Research Centre, School of Mathematics, Statistics and Computer Science, University of KwaZulu-Natal, Durban 4001, South 
Africa}
\author{Adam~D.~Hincks\orcidlink{0000-0003-1690-6678}}\affiliation{David A. Dunlap Department of Astronomy and Astrophysics, University of Toronto, 50 St George Street, Toronto ON, M5S 3H4, Canada}\affiliation{Specola Vaticana (Vatican Observatory), V-00120 Vatican City State}
\author{Arthur Kosowsky}\affiliation{Department of Physics and Astronomy, University of Pittsburgh, Pittsburgh, PA 15260 USA}
\author{Mathew~S.~Madhavacheril}\affiliation{Department of Physics and Astronomy, University of Pennsylvania, 209 South 33rd Street, Philadelphia, PA, USA 19104}
\author{Joshiwa van Marrewijk}
\affiliation{Leiden Observatory, Leiden University, P.O. Box 9513, 2300 RA Leiden, The Netherlands}
\author{Fiona McCarthy}
\affiliation{DAMTP, Centre for Mathematical Sciences, University of Cambridge, Wilberforce Road, Cambridge CB3 OWA, UK}
\affiliation{Kavli Institute for Cosmology Cambridge, Madingley Road, Cambridge CB3 0HA, UK}
\affiliation{Center for Computational Astrophysics, Flatiron Institute, 162 5th Avenue, New York, NY 10010 USA}
\author{Kavilan~Moodley}\affiliation{Astrophysics Research Centre, School of Mathematics, Statistics and Computer Science, University of KwaZulu-Natal, Durban 4001, South 
Africa}
\author{Tony~Mroczkowski\orcidlink{0000-0003-3816-5372}}\affiliation{European Southern Observatory, Karl-Schwarzschild-Str. 2, D-85748, Garching, Germany}
\author{Michael D. Niemack\orcidlink{0000-0001-7125-3580}}\affiliation{
Department of Physics, Cornell University, Ithaca, NY 14853,USA}\affiliation{
Department of Astronomy, Cornell University, Ithaca, NY 14853, USA}
\author{Lyman~A.~Page}\affiliation{Joseph Henry Laboratories of Physics, Jadwin Hall,
Princeton University, Princeton, NJ, USA 08544}
\author{Bruce~Partridge}\affiliation{Department of Physics and Astronomy, Haverford College, Haverford, PA, USA 19041}
\author{Emmanuel~Schaan}\affiliation{SLAC National Accelerator Laboratory 2575 Sand Hill Road Menlo Park, California 94025, USA}\affiliation{Kavli Institute for Particle Astrophysics and Cosmology, 382 Via Pueblo Mall Stanford, CA  94305-4060, USA}
\author{Neelima~Sehgal\orcidlink{0000-0002-9674-4527}}\affiliation{Physics and Astronomy Department, Stony Brook University, Stony Brook, NY USA 11794}
\author{Blake~D.~Sherwin}\affiliation{DAMTP, Centre for Mathematical Sciences, University of Cambridge, Wilberforce Road, Cambridge CB3 OWA, UK}\affiliation{Kavli Institute for Cosmology Cambridge, Madingley Road, Cambridge CB3 0HA, UK}
\author{Crist\'obal Sif\'on\orcidlink{0000-0002-8149-1352}}\affiliation{
Instituto de F\'isica, Pontificia Universidad Cat\'olica de Valpara\'iso, Casilla 4059, Valpara\'iso, Chile}
\author{David~N.~Spergel}\affiliation{Center for Computational Astrophysics, Flatiron Institute, 162 5th Avenue, New York, NY 10010 USA}\affiliation{Department of Astrophysical Sciences, Peyton Hall, 
Princeton University, Princeton, NJ USA 08544}
\author{Suzanne~T.~Staggs}\affiliation{Joseph Henry Laboratories of Physics, Jadwin Hall,
Princeton University, Princeton, NJ, USA 08544}
\author{Eve~M.~Vavagiakis}\affiliation{Department of Physics, Cornell University, Ithaca, NY, USA 14853}
\author{Edward~J.~Wollack}\affiliation{NASA/Goddard Space Flight Center, Greenbelt, MD, USA 20771}

\date{\today}% It is always \today, today,
    % but any date may be explicitly specified

\begin{abstract}
The high electron temperature in galaxy clusters ($>1$\,keV or $>10^7$\,K) leads to corrections at the level of a few percent in their thermal Sunyaev-Zeldovich effect signatures. Both the size and frequency dependence of these corrections, which are known as relativistic temperature corrections, depend upon the temperature of the objects. In this work we exploit this effect to measure the average temperature of a stack of Compton-$y$ selected clusters. Specifically, we apply the ``spectroscopic method" and search for the temperature that best fits the clusters' signal measured at frequencies from 30 to 545 GHz by the Atacama Cosmology Telescope and \textit{Planck} satellite. We measure the average temperature of clusters detected in the ACT maps to be $8.5\pm 2.4$\,keV, with an additional systematic error of comparable amplitude dominated by passband uncertainty. Upcoming surveys, such as the Simons Observatory and CMB-S4, have the potential to dramatically improve upon these measurements and thereby enable precision studies of cluster temperatures with millimeter observations. The key challenge for future observations will be mitigating instrumental systematic effects, which already limit this analysis.
\end{abstract}

\pacs{Valid PACS appear here}
\maketitle

\section{Introduction}\label{sec:intro}

Galaxy clusters are of great cosmological and astrophysical interest.  The hot gas that forms the intra-cluster medium (ICM) radiates and cools, whilst supernova and active galactic nuclei (AGN) heat and expel gas \citep[see, e.g.,][for a review]{Kravtsov_2012}. Together these processes influence the evolution, and fates, of member galaxies. Better understanding these processes would therefore advance our knowledge of galaxy formation and evolution, and elucidate the connection between observable cluster properties and cosmology \citep{Mantz_2016,Mroczkowski_2019,Hlavacek_2022,Moser_2022,Raghunathan_2022}. Cosmologically, differential number counts of these objects as functions of redshift and mass can provide powerful constraints on physics ranging from dark energy to neutrino mass \citep{Holder_2001,Gladders_2007,Louis_2017,Hasselfield_2013,Merloni_2012,Weinberg_2013,planck2014-a30,Mantz_2014,Madhavacheril_2017,Zubeldia_2019,Bocquet_2019,Chaubal_2022,Bleem_2023}. However, to use this information it is necessary to understand the astrophysical processes underpinning the relationships between the observables and the underlying cosmology \citep{Lima_2005,Arnaud_2010,Pratt_2019,Miyatake_2019,Debackere_2021,Robertson_2023,Bocquet_2023}.

The thermal Sunyaev-Zeldovich (tSZ) effect is a spectral distortion to the cosmic microwave background (CMB) that is caused by the inverse Compton scattering of CMB photons off the hot electrons in the intracluster medium \citep{Sunyaev_1972}. The amplitude of the tSZ effect is a direct probe of the electrons' thermal pressure and, especially when combined with measures of the density, can be a powerful probe of these objects' thermodynamics and physical state \citep[e.g.,][]{Mroczkowski_2019}. For cosmology, the tSZ effect provides an almost redshift independent means of detecting galaxy clusters and thereby complements X-ray and optical surveys, which typically probe lower redshifts \citep[see, e.g.,][for a review]{Carlstrom_2002}.

Relativistic corrections describe the percent-level corrections to the tSZ effect, also called the rSZ effect, that arise due to the mildly relativistic nature of electrons in galaxy clusters \citep{Rephaeli_1995,Itoh_1998,Challinor_1998}. Whilst the non-relativistic thermal SZ effect has a spectral signature that is independent of the cluster properties, the specific spectral-shape of the rSZ effect depends sensitively on the galaxy cluster temperature. By inverting this relation, the specific spectral shape of the rSZ signal imprinted by galaxy clusters can be used to infer their electron temperatures.

X-ray measurements are currently the leading method to measure galaxy cluster temperatures. Decades of X-ray measurements have shown that the temperatures of massive galaxy clusters are $\sim 1-10$\,keV \citep[e.g.,][]{Markevitch_1996,Pratt_2009,Reichert_2011,Mantz_2016}. However, X-ray measurements are typically limited to low redshift because X-rays drop off with redshift due to cosmological
surface brightness dimming. The rSZ measurements complement X-ray observations by extending temperature measurements to higher redshift, using the almost redshift-independent properties of the tSZ effect. Measurements of the temperatures of high redshift clusters, when AGN activity peaked, may help determine how feedback processes shaped galaxy clusters \citep[see, e.g.,][]{Lovisari_2022} -- note that this will require the next generation (e.g., Simons Observatory, SPT-3G+ or CMB-S4 \citep{SO_2019,SPT_3G+_2022,S4_2016}) experiments. The temperatures measured by the two probes are also different; SZ temperatures are y-weighted, $\langle n_e T_e\rangle$ where $n_e$ is the electron density and $T_e$ is the electron temperature, whilst X-rays are spectroscopic, $\sim\langle n_e^2 T_e^{\frac{1}{2}}\rangle $\citep[see, e.g.,][for a detailed comparison]{Lee_2020}. Relativistic corrections can also be used as a complementary cosmological probe \citep[e.g.][]{Coulton_2020}. Finally, rSZ measurements offer a means of resolving current calibration uncertainties in X-ray measurements. Chandra and XMM-Newton measurements report temperatures that are $\mathcal O$($10\%$) different, for massive clusters, and this calibration uncertainty is the leading source of error on temperature measurements~\citep{Schellenberger_2015,Wan_2021}. See, e.g., Ref. \citep{DiMascolo_2024} for a detailed discussion of the astrophysical information accessible with rSZ temperature measurements.

In this paper we apply the ``spectroscopic method" developed in Ref. \citep{Remazeilles_2020} to \textit{Planck} NPIPE \citep{planck2020-LVII} and the Atacama Cosmology Telescope (ACT) Sixth Data Release (DR6) maps. This approach makes maps of the sky that show the difference between the electron temperature, $T_e$, and a trial temperature, $\bar{T}_e$. By iterating through trial temperatures we can find a point that nulls the signal, which indicates the gas temperature at that location matches the trial temperature. By combining \textit{Planck}'s full-sky, broad frequency ($100-545$\,GHz) maps with ACT high-resolution data sets, we can disentangle the small relativistic signal from other sky contaminants. Several past works have used the SZ effect to constrain electron temperatures \citep[e.g.,][]{Prokhorov_2012,Hurier_2016,Erler_2018,Remazeilles_2024}. The current state-of-the-art is Ref.~\citep{Erler_2018} that constrained the average temperature of \textit{Planck} clusters to be $4.4_{-2.0}^{+2.1}$\,keV, Ref. ~\citep{Remazeilles_2024} that reanalyzed the \textit{Planck} clusters to find an average temperature of $4.9\pm2.6$\,keV, Ref. \citep{Perrott_2024} assessed the impact that neglecting the rSZ corrections had on the \textit{Planck} SZ scaling relations, and Ref. \citep{Butler_2022} that used Bolocam data to constrain the temperature of the cluster RX J1347.5-1145. The high-resolution ACT data used in this work, which improves upon \textit{Planck}'s resolution at $143$\,GHz by a factor of $\sim 5$, allows a better characterization of the SZ effects below $\sim 220$\,GHz.

The paper is structured as follows: in \cref{sec:stacking} we describe our method to isolate the relativistic contributions to the SZ effect and we overview the data sets in \cref{sec:data}. In \cref{sec:sims}, we demonstrate and validate our approach on realistic, non-Gaussian simulations. We present the measurement in \cref{sec:results} and constraints on the temperature-Compton-$y$ relation in \cref{sec:temp_evolution}. Finally we present our conclusions and outlook in \cref{sec:conclusions}.

\section{Spectroscopic Methodology }\label{sec:stacking}

\begin{figure}
    \centering
\includegraphics[width=.47\textwidth]{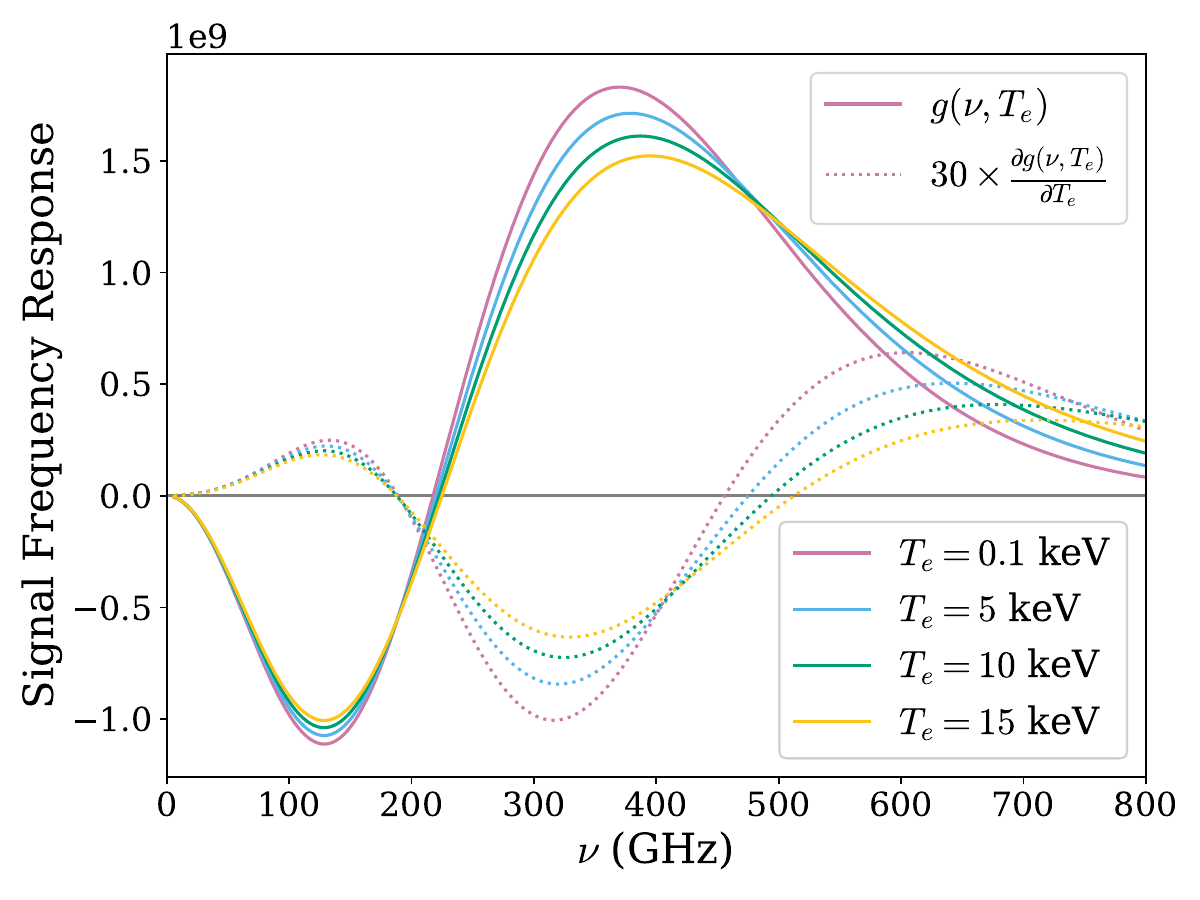}
    \caption{The frequency dependence of the relativistic thermal SZ signal (solid, in Jy/Sr) and its derivative with respect to electron temperature (dotted, in Jy/Sr/keV) for a range of different electron temperatures. This analysis utilizes the changes in the spectral response of the SZ signal to constrain the temperature of galaxy clusters.}
    \label{fig:responses} 
\end{figure}

Non-relativistic electrons are those that have temperatures, $T_e$, satisfying  $T_e k_b\ll m_e c^2$, where $k_B$ is the Boltzmann constant, $m_e$ is the electron mass and $c$ is the speed of light. Cosmic non-relativistic electrons induce thermal Sunyaev-Zeldovich anisotropies in the CMB $\Delta I$, in direction $\mathbf{n}$ and at observation frequency $\nu$, as
\begin{align}\label{eq:non-relSZ}
\Delta I^\mathrm{non-rel}(\mathbf{n},\nu) = g^\mathrm{non-rel}\left(\nu\right) y(\mathbf{n}),
\end{align}
where $g^\mathrm{non-rel}\left(\nu\right)$ is the frequency response function from inverse-Compton scattering off non-relativistic, thermal electrons and $y(\mathbf{n})$ is the Compton-$y$ parameter. The Compton-$y$ parameter traces the integrated electron pressure as
\begin{align}
    y(\mathbf{n}) = \frac{\sigma_T}{m_ec^2}\int\limits_{los} \mathrm{d}l\,P_e(\mathbf{n},l),
\end{align}
where $\sigma_T$ is the Thomson cross section, and $P_e$ is the thermal electron pressure.
For non-relativistic electrons the frequency response function has a well-known form
\begin{align}\label{eq:non-rel-SZ-freq}
&g^\mathrm{non-rel}(\nu)   = \left.\frac{\mathrm{d}B(\nu,T)}{\mathrm{d}T}\right\vert_{T= T_\mathrm{CMB}} \left[ x \frac{e^{x} + 1}{e^{x} -1} - 4 \right],
\end{align} 
where  ${\mathrm{d}B(\nu,T)}/{\mathrm{d}T}$ is the derivative of the CMB blackbody,  $x=\frac{h \nu}{k_B T_\mathrm{CMB}}$, $T_\mathrm{CMB}$ is the CMB temperature and $h$ is Planck's constant. As visible in \cref{fig:responses}, this frequency function has a null at 217\,GHz.

Electrons in massive clusters can have temperatures $\gtrsim 10$\,keV and so the non-relativistic classification is no longer accurate. Instead, the induced anisotropies are given by
\begin{align}\label{eq:rSZ}
\Delta I(\mathbf{n},T_e,\nu) = g\left(\nu,T_e(\mathbf{n})\right) y(\mathbf{n}),
\end{align}
where the frequency response function $g\left(\nu,T_e(\mathbf{n})\right)$ now depends on the electron temperature $T_e$. The non-relativistic signal is the limit $T_e\ll 0.1\,$keV, and the relativistic response approaches \cref{eq:non-rel-SZ-freq} in this limit. As shown in \cref{fig:responses}, the shape of the spectral response changes with temperature, with changes at the level of a few percent occurring at $5$\,keV at $150$\,GHz.   Unlike the non-relativistic case,  an analytic form is not available\footnote{See \citep{Challinor_1998,Itoh_1998,Nozawa_2006} for analytic fits to the corrections, which require higher order polynomials and are valid for a range of frequencies and temperatures} for a general temperature and so we compute these signals using \textsc{szpack}, described in Refs.~\citep{chluba_2012,chluba_2013}. We define relativistic corrections as $\Delta I(\mathbf{n},T_e,\nu)-\Delta I^\mathrm{non-rel}(\mathbf{n},\nu)$, 
i.e., the difference between the relativistic tSZ anisotropies and those computed with the non-relativistic response.

Our approach is based on the method presented in Ref.~\citep{Remazeilles_2020}. The first step of this approach is to consider a Taylor expansion of the full response, given in \cref{eq:rSZ}, about a trial temperature, $\bar{T}_e$. With this expansion we can rewrite \cref{eq:rSZ} as 
\begin{align}\label{eq:rsz_taylor}
& \Delta I(\mathbf{n},T_e,\nu)=\nonumber \\ &y(\mathbf{n}) \left[ g^{(0)}\left(\nu,\bar{T}_e\right)  +  g^{(1)}\left(\nu,\bar{T}_e\right)\left(T_e(\mathbf{n})-\bar{T}_e \right)  \right] \nonumber  \\ & + \mathcal{O}\left[(T_e(\mathbf{n})-\bar{T}_e)^2\right],
\end{align}
where $g^{(0)} = g|_{T_e=\bar{T}_e}$ is the spectrum of gas at temperature $\bar{T}_e$  and $g^{(1)}= \frac{\partial g}{\partial T_e}|_{T_e=\bar{T}_e}$ is the spectral shape of the first order correction. We then use the internal linear combination (ILC) method to make a map of the sky that isolates signals with a spectrum given by $g^{(1)}$; we call this map the $g^{(1)}$ map.  Next, we examine the signal at the location of galaxy clusters. If the signal is positive (negative), then the true galaxy  cluster temperature is greater (less) than $\bar{T}_e$. This process can be iterated to find a $\bar{T}_e$ for which the signal vanishes. This $\bar{T}_e$ is then the $y$-weighted cluster temperature; see, e.g., Ref.~\citep{Lee_2022} for a discussion of different temperature measures. In this work we use 38 trial temperatures between $0.1$\,keV and $35\,$keV. The trial temperatures are more densely sampled for $\bar{T}_e<10\,$keV, the region of expected cluster temperatures. With very low noise measurements this can be performed cluster by cluster, as suggested in Ref.~\citep{Remazeilles_2020}. In this work we are not able to measure individual cluster temperatures and instead apply this to stacks of objects.

\subsection{Contamination from radio sources}
Active galactic nuclei produce signals in millimeter observations through synchrotron emission. These radio sources are mostly unresolved point sources in CMB observations. The frequency dependence of their emission is approximately power law, $I(\nu)\propto \nu^\alpha$, where the spectral index, $\alpha$, depends on the source with typical values $-0.8\lesssim \alpha\lesssim -0.1$. Emission from radio sources can ``infill" the tSZ decrement and
thereby bias studies of the tSZ effect \citep{Dicker_2021,Dicker_2024}. We exclude clusters with bright radio sources and a detailed examination of their impact is given in \cref{sec:contaminants}. 

\subsection{Contamination from dusty star forming galaxies}
A second key contaminant is thermal emission from dust in star forming galaxies. In the galaxies' rest frame, this emission is in the infrared and it is redshifted to the millimeter. Most dusty star forming galaxies are too faint to be observed individually and are unresolved in CMB datasets. Instead the emission from many galaxies together forms the diffuse cosmic infrared background (CIB). Whilst the CIB is primarily sourced from high-redshift, low mass objects, some CIB galaxies are found in groups and clusters. There is also the lensing of the dusty star forming galaxies by cluster member galaxies. The CIB emission is strongest at high frequencies (see \cref{eq:greyBody}), where the rSZ signal responds most strongly to changes in temperature. Whilst the SEDs of the CIB and rSZ are different, the CIB is much brighter and, in the noisy measurements at sparse frequency intervals used in this analysis, can significantly bias rSZ inferences.

We use the constrained needlet internal
linear combination (NILC) method, developed in Ref.~\citep{Remazeilles_2011b}, to mitigate biases from both the CIB and the non-relativistic tSZ effect. The constrained NILC removes contaminating sky signals with known spectral signatures. This is achieved by using Lagrange multiplier constraints to ensure the NILC map has no contributions from components with a specific frequency dependence. There is an increase in noise for each component removed, i.e. each additional constraint. Applying a constraint is also called ``deprojecting" a component and we use these terms interchangeably. First, we deproject sky signals with the $g^{(0)}(\bar{T}_e)$ spectrum. This prevents the zeroth order component of the Taylor expansion, which is $\mathcal O(10 \times)$ larger, from biasing our measurement. Second, we deproject a component consistent with the spectrum of the cosmic infrared background. This mitigates contamination from dusty star forming galaxies. 

The ``true" CIB spectrum is unknown and likely spatially varying \citep{Bethermin_2011,Mak_2017}. A leading order approximation can be achieved with a modified blackbody (MBB) spectrum, defined as 
\begin{align}\label{eq:greyBody}
I^\mathrm{MBB}(\nu) = \frac{A \left(\frac{\nu}{\nu_0}\right)^{3+\beta}}{\exp\frac{h\nu}{k_B T_\mathrm{CIB}}-1}
 \end{align}
where $A$ is a normalization constant, $v_0=350\,$GHz is a reference frequency, $\beta=1.7$ and $T_\mathrm{CIB}=10.7$\,K, as obtained to fits to \textit{Planck} CIB measurements in Ref. \citep{McCarthy_2023a}. Applying a constraint to deproject the modified blackbody suppresses CIB contamination. Refs. \citep{Chluba_2017,Azzoni_2021} demonstrated that Taylor expanding around an approximate spectral energy distribution and deprojecting the terms of this expansion, known as the moment expansion, can account for our imperfect knowledge of the ``true" spectrum and/or spatial variation of the spectrum. Exploiting these results, we mitigate any residual CIB contamination through an additional constraint that deprojects the derivative of the MBB with respect to spectral index, $\beta$. See, e.g.,\ Ref. \citep{McCarthy_2023a,Coulton_2023} for a detailed discussion of CIB mitigation using the moment expansion. In \cref{sec:contaminants} we explore the impact of different CIB modelling assumptions. We refer the reader to Ref. \citep{Coulton_2023} for a detailed description of our implementation of the NILC method. The NILC maps produced for this work have a $1.6^\prime$ full-width-half-maximum (FWHM) Gaussian beam applied to them.

\subsection{Contamination from the kinetic Sunyaev-Zeldovich effect}
Galaxy clusters also produce a kinetic Sunyaev-Zeldovich (kSZ) signal in millimeter observations \citep{Sunyaev_1980}. For a typical cluster, the kSZ signal is $\sim 3\times$ larger than the rSZ signal. The kSZ anisotropies have the same spectrum as the primary CMB anisotropies, temperature distortions to a blackbody at $2.725$\,K \citep{Fixsen_2009}. We do not deproject the CMB signal and so the kSZ will be present in our NILC maps. Note that the kSZ signal will likely be reduced by the NILC method as the minimum variance signal map will typically have reduced contaminant signals compared to the input maps. The kSZ effect can be either positive or negative, depending on whether the line-of-sight component of the cluster's velocity is towards or away from the observer. In this work we consider stacks of clusters so the kSZ should average to zero. Specifically, we include $>500$ objects per stack to ensure that the residual kSZ is small ($<10\%$ of the expected rSZ signal). Whilst the number of objects was determined from simulations it can be heuristically understood by considering the rSZ and kSZ signals. The kSZ signal depends on $n_e v/c$ (where $v$ is the line-of-sight velocity) and the tSZ depends on $n_e T_e$. Cluster velocities are ${\mathcal O}(v/c)\sim(10^{-3})$ whilst temperatures are ${\mathcal O}(k T_{\rm e} / (m_{\rm e} c^2)) \sim (10^{-2})$. The rSZ effect is a $3-5\%$ correction on the tSZ effect. Stacking 500 objects whose kSZ signal fluctuates randomly around zero, like noise, reduces the stacked kSZ to  ${\mathcal O}(10^{-3}/\sqrt{500})$; thus, the kSZ is expected to be $\sim10\%$ of the rSZ signal.

\subsection{Mitigation of large-scale noise}
The NILC maps exhibit a significant amount of large-scale noise. This noise arises from residual Galactic foregrounds, which have significant power on large scales, and instrumental noise. To remove this noise, we high-pass filter the ILC maps, keeping modes with $\ell\geq 500$. Whilst a filter keeping $\ell \geq 200$ already mitigates most of the large-scale noise, $\ell=500$ was chosen as it approximately corresponds to the fundamental mode of the cutouts that we use for each cluster. Modes larger than this would add to a mean mode of the patch and so removing these additional modes reduces the covariance of our measurements. We also filter out small-scale modes, those with $\ell>3200$. This limit is set by instrumental systematic effects in ACT data, discussed in \cref{sec:dataConsistency}, and by the limited number of low-noise observations at these scales. The NILC method needs a minimum of four frequency channels (one for the component of interest and three for the constraints). We use three ACT channels between 90 and 220\,GHz and eight \textit{Planck} channels between 30 and 545\,GHz. On small scales ($\ell \gtrsim 3200$) none of the \textit{Planck} channels provide significant information and this impacts the maximum $\ell$ we can use. Finally note that as this method uses the frequency dependence to measure the temperature, these filtering steps do not bias the temperature measurements.

\subsection{Overview of the analysis pipeline}\label{sec:pipelineOverview}
Bringing these pieces together, our pipeline is as follows: first we use the NILC algorithm to produce a map of the $g^{(1)}$ component at trial temperature, $\bar{T}_e$. In the NILC, we use two constraints to mitigate CIB contamination. We then filter the map to remove the largest and smallest scale modes. We then extract $20^\prime\times20^\prime$ cutouts around the location of clusters and average them. We extract cutouts by projecting onto a tangent plane of equal area pixels, as in \citep{Schaan_2021,Coulton_2024}; this step accounts for the curvature of the sky and the unequal area of pixels in the Plate Carr\'ee projection used for the NILC map. We compute 1D radial profiles by azimuthally-averaging the 2D cutouts. This procedure exploits the approximate rotational symmetry of the signal to average down the noise. 

We compute the covariance of our measurements through repeating our procedure at random locations in the map far from detected clusters and point sources (see \cref{sec:data} for more details). We extract 5,000 1D profiles from these random locations and compute the covariance of the data points. As the signal in the $g^{(1)}$ map is proportional to Compton-$y$, at the random locations away from clusters we just measure noise. By assuming that each cutout is independent, we obtain the covariance matrix of our mean profile by rescaling the covariance matrix by the number of objects in our catalog. The resulting covariance matrix is shown in \cref{fig:cor_mat}. The high pass filtering, discussed above, ensures that the patches are approximately independent. We verified that this procedure matches a covariance matrix obtained from bootstrapping. For the bootstrap method, we create a new data vector by drawing, with repetition, a matching number of objects from our cluster catalog and computing the mean of the 1D profiles for the new catalog. We repeat this 100,000 times and measure the covariance. The two methods of estimating the covariance matrix agree at the $\sim1\%$ level.

We assume that the data are Gaussian distributed, an accurate approximation as the maps are noise dominated, and use a Gaussian likelihood, $\mathcal{L}(\mathbf{d}|\bar{T}_e)$, for our data. We choose an uninformative prior that has support between $0\,$keV$<\bar{T}_e<35\,$keV and sample the following posterior
\begin{align}\label{eq:likelihood}
    \ln P(\bar{T}_e|d) \propto -\mathbf{d}C^{-1}\mathbf{d},
\end{align}
where $\mathbf{d}$ is the 1D radial measurements at a given $\bar{T}_e$ and the covariance matrix, $C$, is computed for each trial temperature. This posterior is maximized when the trial temperature matches the true temperature, as the signal vanishes when $\bar{T}_e=T_e$. The structure of our prior, and the reasoning for this choice, is discussed in \cref{app:prior_choice}.
For high-resolution, low noise measurements the temperature could be fit for each radial bin of the 1D profile to constrain the temperature profile. Given the limited statistical power of our data we analyse all the radial bins simultaneously. This would be optimal if clusters were isothermal. Whilst real clusters are not isothermal \citep{Vikhlinin_2005,Leccardi_2008,Zhu_2016,Lee_2020,Lee_2022}, we are not sensitive to these differences at our current resolution and noise levels.

\begin{figure}
    \centering
\includegraphics[width=.48\textwidth]{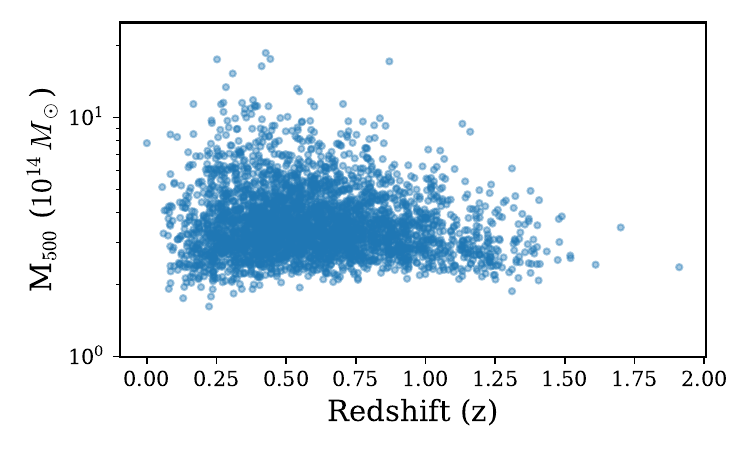}
    \caption{The distribution of masses and redshifts for 3495 of the clusters used in this work. The median mass is $3.4\times 10^{14}\,\rm M_\odot$ and the median redshift is $0.56$. Note that this plot only includes clusters with optical redshifts. Redshifts are not necessary for this measurement and so we additionally include 1195 clusters that are currently without redshifts. The total number of clusters used is 4690. This cluster catalog is prepared in a similar manner to \citep{Hilton_2021} and we refer the reader there for more details. }
    \label{fig:cluster_mass_and_redshift} 
\end{figure}
\section{Data Sets}\label{sec:data}
\subsection{Input frequency maps}
In this work we combine data from the \textit{Planck} satellite and ACT. The \textit{Planck} experiment's broad frequency coverage (30\,GHz-545\,GHz) and precision large-scale measurements are complementary to ACT's low-noise, high-resolution measurements.

The \textit{Planck} satellite observed the full sky from  2009 to 2013 at 9 frequencies. We use data from the NPIPE \textit{Planck} release \citep{Planck_int_LVII} at the 8 frequencies between $30$\,GHz and $545$\,GHz. The FWHM of the \textit{Planck} beam ranges from $32^\prime$ at $30$\,GHz to a FWHM$=4.67^\prime$ at $545$\,GHz. \textit{Planck}'s high frequency channels are essential for removing contaminating emission from the dusty star forming galaxies that make up the CIB. 

We use ACT \citep{Ho_2017,Choi_2018, Henderson_2016} data from the Data Release 6 (DR6). The data used here cover observations between 2017 and 2022 and three frequencies: f090, f150 and f220. The ACT frequency names are defined in Ref. \citep{Naess_2020} and have central frequencies at approximately $98\,$GHz, $150\,$GHz and $224$\,GHz, respectively. The ACT beam FWHM at f150 is five times smaller than \textit{Planck}'s beam at $143$\,GHz: $1.4^\prime$ compared to $7.22^\prime$. The data are processed as in \citep{Coulton_2023}, but with some improvements. We use an updated version of the map-making pipeline that reduces systematic effects. Whilst the differences between the maps are very small, given the small size of the rSZ signal it is prudent to use the most accurately characterized maps. We also exclude the Polarized Array 4 (PA4) f150 data due to a currently unresolved calibration issue. These maps will be presented in an upcoming paper

\begin{figure*}
  \centering
  \begin{subfloat}[Non-relativistic  SZ Simulations\label{fig:sims_SZ}]{\includegraphics[width=.47\textwidth]{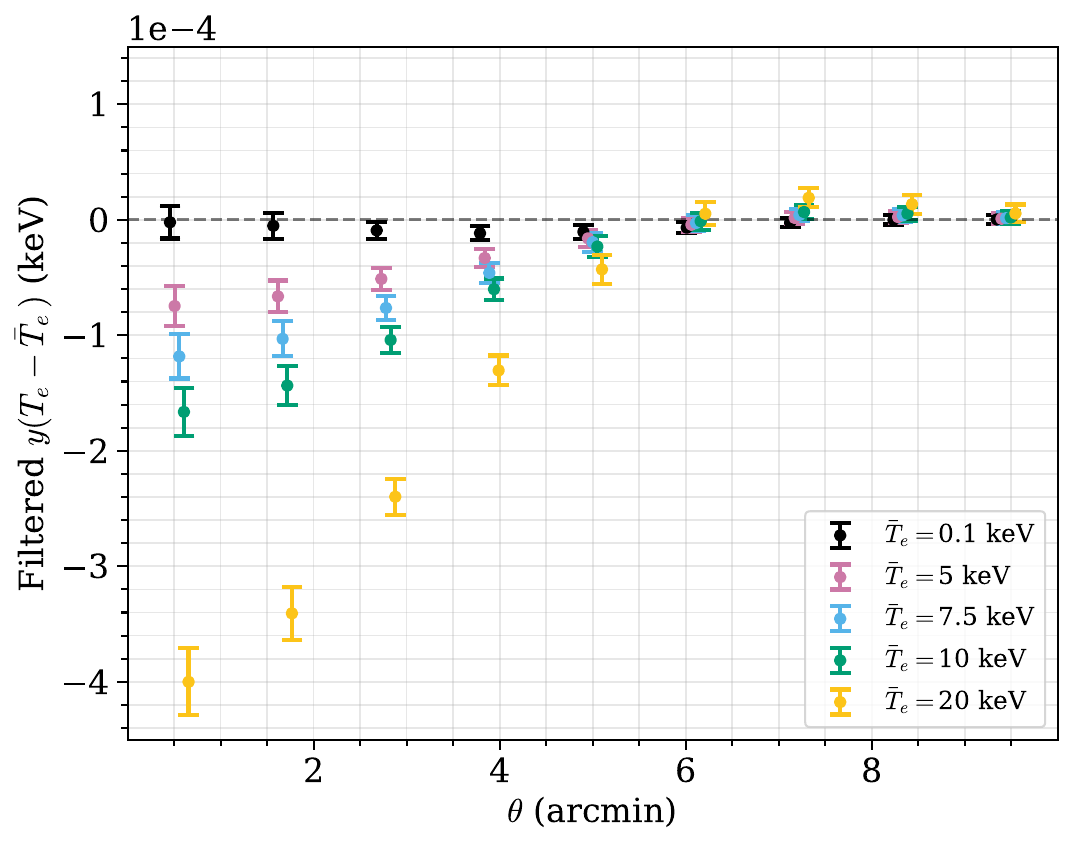} }
  \end{subfloat}
  \hfill
  \begin{subfloat}[Relativistic  SZ Simulations\label{fig:sims_rSZ}]{\includegraphics[width=.47\textwidth]{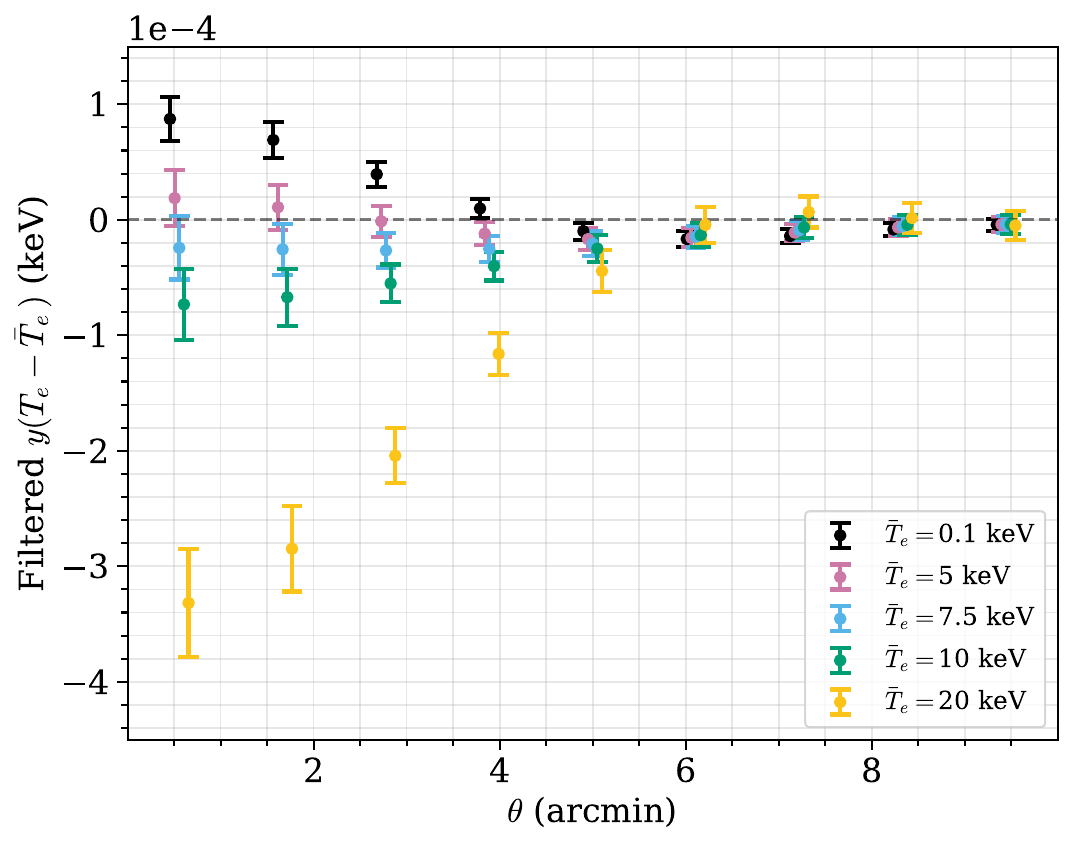} }
  \end{subfloat}
\caption{A demonstration of the ``spectroscopic" method used in this work on realistic, non-Gaussian ACT \& \textit{Planck} simulations. We show the angle average measurements of the g$^{(1)}$ maps, \cref{eq:rsz_taylor}, at cluster locations for five trial temperatures. The g$^{(1)}$ maps show the Compton-$y$ weighted difference between the cluster temperature and trial temperature. The sky simulations combined the extragalactic Sehgal et al. simulations \citep{Sehgal_2010} with \textsc{pysm} \citep{Thorne_2017,Zonca_2021} Galactic models. We test two versions of the Sehgal et al. simulations: one containing only the non-relativistic SZ effect (left) and one containing the relativistic SZ effect. We apply the method to a stack of $2921$ clusters. The method is designed to return a null signal for estimated cluster temperatures that match the simulated temperatures. It works as expected, for simulated mean temperatures of T$=0$\,keV (left) and T$=5.6$\,keV (right).  This demonstrates the sensitivity of our method to cluster temperatures and robustness to contaminant sky signals. We note that the ringing feature, as seen in the yellow data points, arises from the harmonic space filtering used in this analysis.
}
\label{fig:sehgal_sims}
\end{figure*}
As in Ref. \citep{Coulton_2023}, we inpaint the region around very bright point sources, those with signal-to-noise ratios (SNRs) $\geq70 $; subtract all remaining point sources that are
detected at SNR\,$\geq 5$ in either ACT or \textit{Planck} data, and mask the dusty regions around the Galaxy, primarily with the \textit{Planck} 70\% sky mask. The output NILC maps have a $1.6^\prime$ FWHM Gaussian beam that we do not deconvolve in this analysis.
\subsection{Cluster catalog}
We use the revised ACT cluster catalog, obtained by applying the methods from Ref. \citep{Hilton_2021} to the latest ACT DR6 maps. The mass and redshift distribution is shown in \cref{fig:cluster_mass_and_redshift}, and this catalog will be described in detail in a future
publication. We use all clusters with SNR~$\geq 5\,$. Clusters located within 2$^\prime$ of subtracted point sources are removed from the catalog. This avoids any potential bias from imperfectly subtracted point sources and was also used in Ref. \citep{Li_2021}. Similarly we avoid any cluster within 10$^\prime$ of a masked or inpainted pixel. Finally, we exclude clusters that may be strongly contaminated by radio sources. We identify strongly contaminated clusters as those with a total radio flux within a 2.5$^\prime$ disc exceeding 50\,mJy at 1367.5\,MHz.  This removes 28\% of objects from the catalog.\footnote{This is larger than the 13\% of clusters that Ref. \citep{Dicker_2021} found to have fluxes contaminated by more than 5\%. The dramatically different frequencies of RACS and the objects studied in  Ref. \citep{Dicker_2021} mean that we cannot map this removal fraction to a contamination flux level from Ref. \citep{Dicker_2021}.} The radius of the disc is chosen to match the angular size of a typical ACT cluster, so will mitigate the impact of correlated radio sources. We use the Rapid Australian SKA Pathfinder (ASKAP) Continuum Survey (RACS)-mid catalogs from Ref. \citep{Duchesne_2023a} to assess this criterion. The RACS catalogs are ideal for this purpose as they cover the entire ACT footprint and are 95\% complete to 1.6\,mJy. This leaves 4690 clusters, of which 3495 have redshift estimates.

 A disadvantage of the RACS catalog is that it is at a much lower frequency than our observations. Unfortunately there is not an equivalently deep catalog at higher frequencies. We did also examined a higher frequency (20\,GHz) survey, AT20G \citep{Murphy_2010}. We found that this catalog has seven clusters with nearby sources that are not found in the RACS survey.  However, the lack of depth (40 mJy/beam) means that there are $\mathcal{O}$(100s) of clusters with sources in the RACS catalog that are not in the AT20G catalog. As such we conclude that the lower frequency RACS catalog is a better catalog to test and mitigate radio contamination. We discuss any residual contamination of our rSZ signal by embedded radio source in more depth in \cref{sec:contaminants}.

\begin{figure}
    \centering
\includegraphics[width=.47\textwidth]{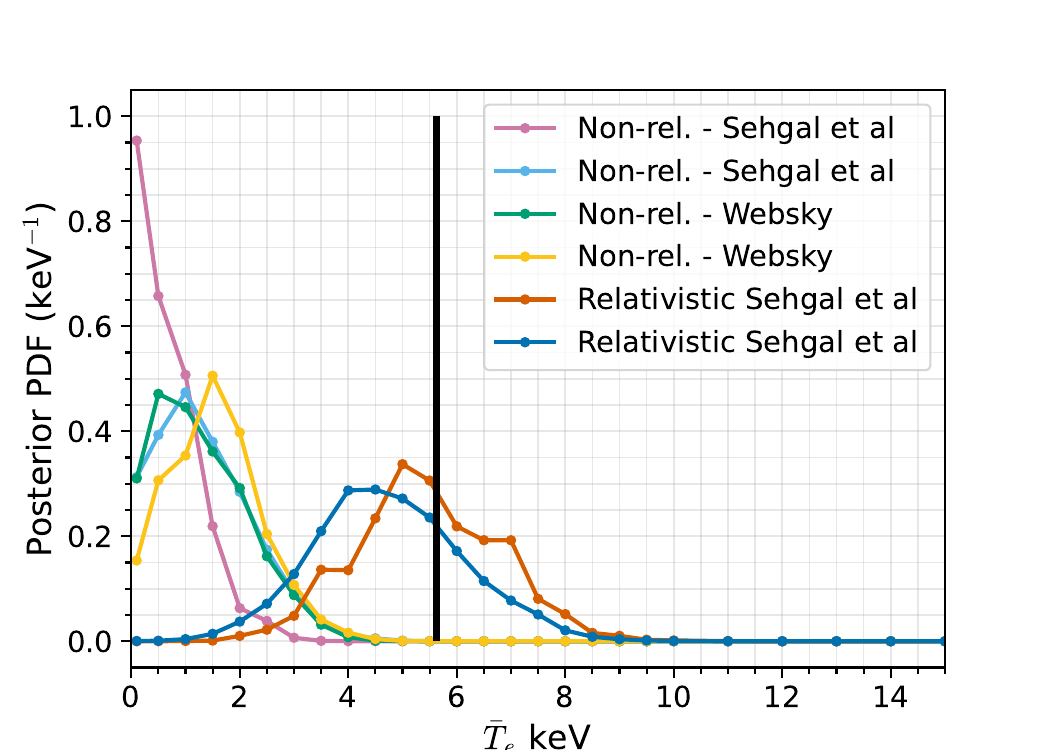}
    \caption{Posterior distributions for the average temperature measured via the relativistic SZ signal. We analyze stacks of clusters in six different simulations: two realizations of the sky from the relativistic SZ Sehgal et al. simulations (orange and blue), two from the non-relativistic SZ Sehgal et al. simulations (pink and cyan) and two from the non-relativistic  SZ Websky simulations (green and yellow). Each stack has around $3000$ clusters.  This estimator is
consistent with zero for the simulations that do not include relativistic corrections. This demonstrates that our method is not biased by residual foregrounds. On the other hand, our method correctly recovers the input, non-zero temperature (vertical black line) for the relativistic simulations.
  }
    \label{fig:sim_posteriors} 
\end{figure}

\section{Validation of methodology}\label{sec:sims}

To validate our approach we use a set of simulated ACT DR6 \& \textit{Planck} NPIPE observations. These simulations are described in detail in Ref. \citep{Coulton_2023} and so here we only summarize their key properties. 

The simulations combine realistic noise, generated from the tiled method presented in Ref. \citep{Atkins_2023} and end-to-end \textit{Planck} simulations \citep{planck2014-a14,Planck_int_LVII}, with state-of-the-art non-Gaussian sky simulations. The Galactic sky components are obtained from the \textsc{pysm} sky model \citep{Thorne_2017,Zonca_2021}.\footnote{ We use models ``d1", ``s1", ``a1" and ``f1".} The extra-galactic components come from two simulation suites: the Sehgal et al. simulations \citep{Sehgal_2010} and the Websky simulations \citep{Stein_2018,Stein_2020}. There are two versions of the Sehgal et al. simulations: one that includes the SZ effects (both tSZ and kSZ) using the non-relativistic approximations and one that includes relativistic corrections for both the kSZ and tSZ, using the 4$^\mathrm{th}$ order perturbative model of Ref. \citep{Nozawa_1998}. The Websky simulations complement the Sehgal et al. simulations through the use of different models for the radio galaxies, CIB, and the kinetic and thermal SZ effects. However, the Websky simulations do not include the relativistic SZ effect. Both suites of non-relativistic simulations enable an assessment of the level of any foreground bias in the absence of any signal. The relativistic Sehgal et al.\ simulations are used to test whether our approach can recover an input relativistic signal correctly. For each of the extra-galactic simulations, we extract two almost independent realizations of the ACT footprint.  The Websky simulations include the \textit{Planck} $30-545$\,GHz observations and the ACT f090, f150 and f220 observations, each with the appropriate instrumental beams \citep{planck2013-p03c,Lungu_2022}. No 545\,GHz relativistic maps are available for the Sehgal et al.\ simulations so we exclude that map from the relativistic analysis. Given the reduced frequency range, we are forced to exclude the derivative of the CIB SED constraint from the NILC when analyzing the Sehgal et al.\ simulations. This means that we assume that the CIB can be effectively removed with a single modified black body. We do not run point source and cluster finders on these simulations. Instead we construct point source masks based on the input component maps, masking radio sources with flux greater than $9$\,mJy at 90\,GHz  and dusty star forming galaxies with  flux greater than $30$\,mJy at 220\,GHz. These flux cuts approximately match those used on the ACT data. The cluster catalogs used in our analysis consist of all massive halos from the Websky catalog with $M_\mathrm{vir}>4.2\times 10^{14}\,{\rm M}_\odot$ and that are not excluded by the masks. This mass cut approximately matches the ACT cluster sample (note that \cref{fig:cluster_mass_and_redshift} shows $M_{500}$ not $M_\mathrm{vir}$ and these quantities differ by $\sim 60\%$).
 
In \cref{fig:sehgal_sims}, we demonstrate the application of our method to one realization of the Sehgal et al.\ sky simulations. \cref{fig:sims_SZ} shows a set of measurements on simulations with the non-relativistic SZ effects. For all trial temperatures greater than $0.1$\,keV, we see a negative signal. This implies the temperature of these is $<0.1$\,keV, i.e.\ non-relativistic. \cref{fig:sims_rSZ} shows the equivalent result for the relativistic SZ simulations. In this case we see a positive signal for low trial temperatures (black), a null at $\sim 5$\,keV and then a negative signal for greater trial temperatures. It is this feature that we use to constrain the temperature of galaxy clusters.

\begin{figure}
    \centering
    \includegraphics[width=.47\textwidth]{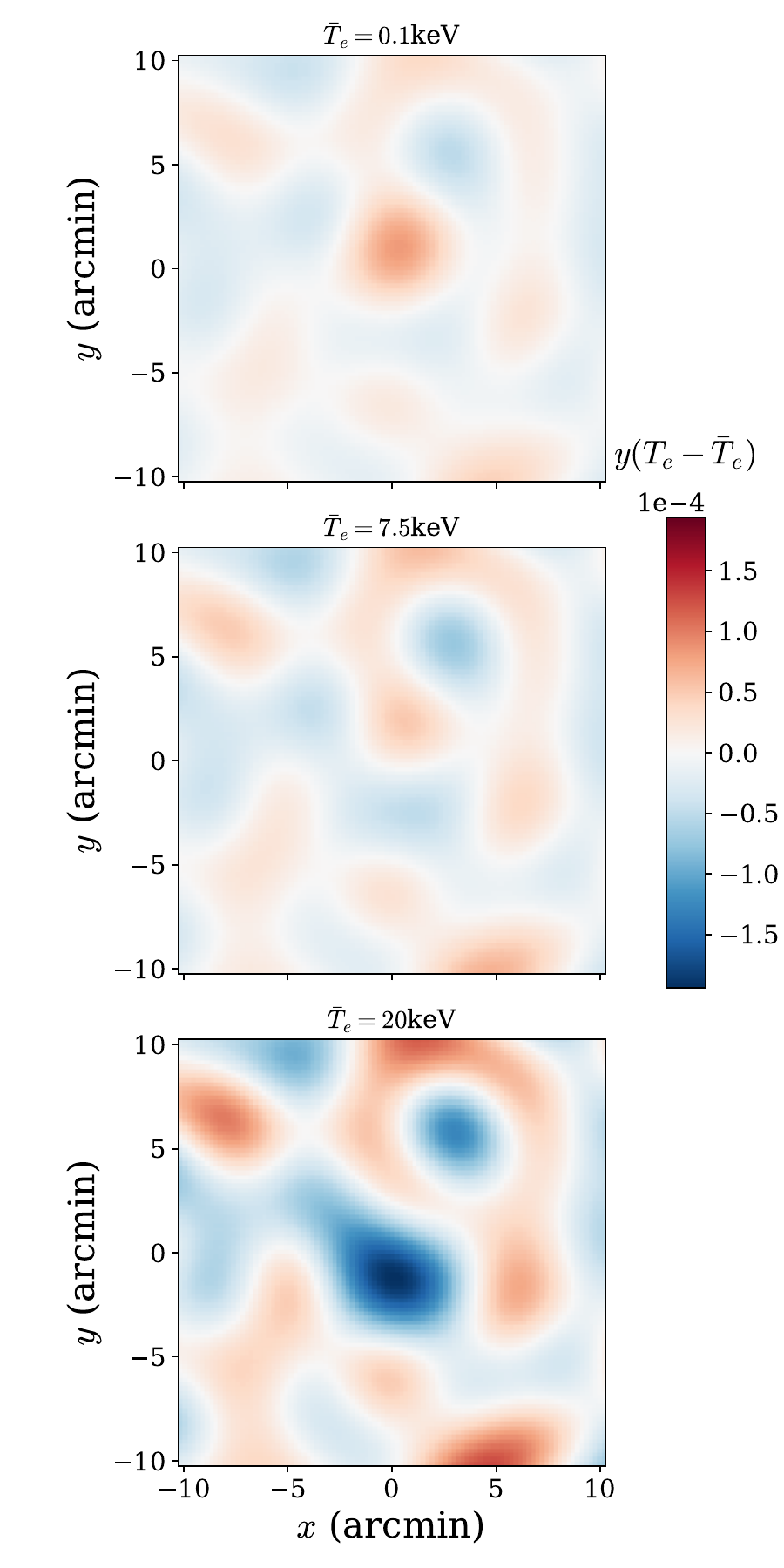}
    \caption{2D stacks of filtered ACT \& \textit{Planck} data at the location of ACT Compton-$y$ selected clusters with SNR$>5$. We show the stacks of the $g^{(1)}$ maps, which measure the Compton-$y$ weighted difference between the cluster temperature and the trial temperature as given by \cref{eq:rsz_taylor}, for three trial temperatures. A null signal is expected for a trial temperature that matches the true mean cluster temperature. The positive signal seen at low trial temperatures, $\bar{T}_e$, indicates a non-zero temperature for our cluster sample. Likewise, the negative signal seen at higher trial temperatures rules out extremely hot clusters.}
    \label{fig:2dStacks}
\end{figure}
We sample the temperature posterior, \cref{eq:likelihood}, with the results shown in \cref{fig:sim_posteriors}. Here we see that the temperature posteriors for the four realizations of the non-relativistic simulations (two from Sehgal et al. and two from Websky) are consistent with zero. This implies our method is not biased by the other non-Gaussian sky components, at least as implemented in these simulations. This is interesting as we do not apply the derivative constraint to the Sehgal et al. simulations. This implies that simply deprojecting the CIB is sufficient for these noise levels and for this simulation.\footnote{Given the challenges of simulating the CIB, this conclusion may not be true for other CIB modelling choices and preliminary tests show that this condition is necessary for the Websky simulations.} Given the uncertainties in the ``true" CIB we retain this constraint when analyzing the data.
Next, we see that the temperature posteriors of the two simulations with a relativistic SZ signal peak away from zero and exclude zero at $4.5$ and $3.9\,\sigma$. We find that the posteriors are consistent with the expected y-weighted temperature signal of $5.6$\,keV. This expected temperature is computed by fitting a temperature to the spectrum of each simulated cluster. The spectrum is given by the simulated relativistic SZ fluxes, averaged in $r_{500}$, at frequencies between 30\,GHz and 350\,GHz.\footnote{We do not use the central temperature given in the Sehgal et al. catalogs as the temperature of the clusters changes rapidly with radius \citep{Bode_2003,Ostriker_2005,Bode_2007}. This is can be seen in, e.g., Fig 2 of \citep{Ostriker_2005}. This rapid radial change means that the temperature of the region that sources the bulk of the SZ effect will be different from the central temperature.}
 
\section{Measurement of the average temperature}\label{sec:results}
\begin{figure}
    \centering
    \includegraphics[width=.47\textwidth]{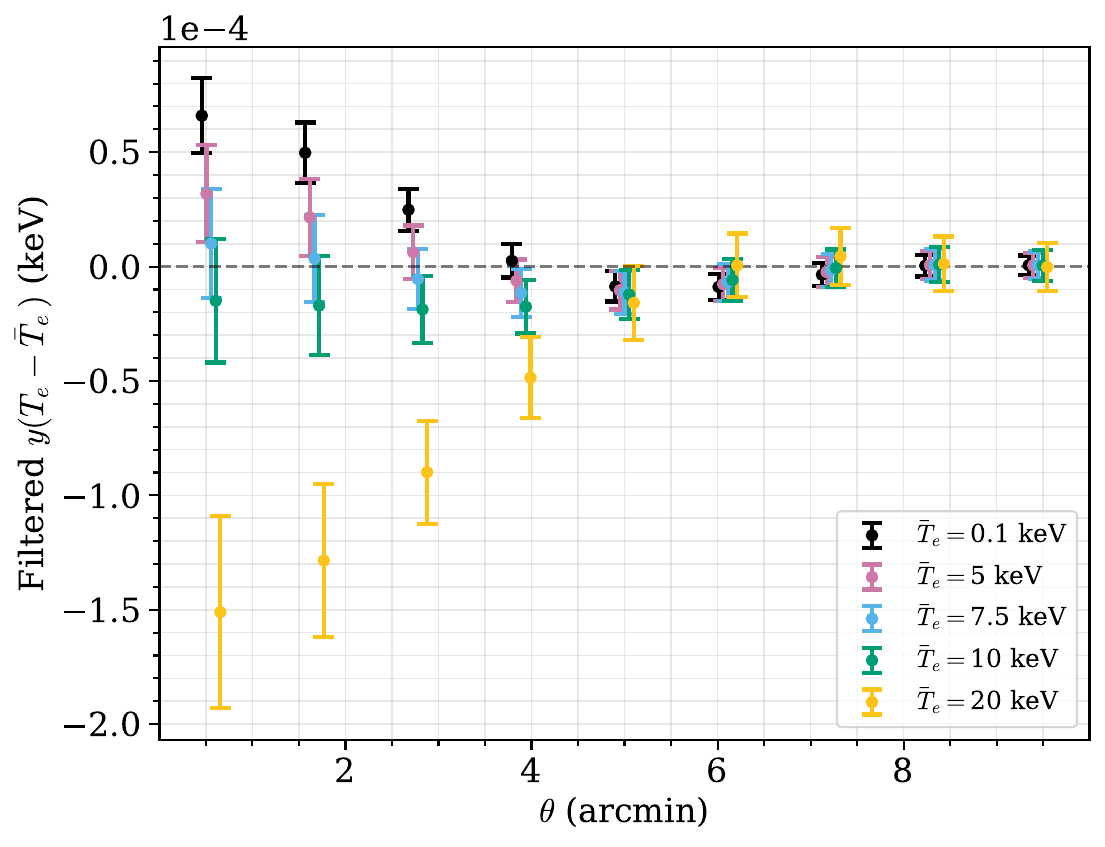}
    \caption{1D profiles of the spectroscopic measurements for all ACT Compton-$y$ selected clusters with SNR$>5$. This is obtained as an azimuthal average of \cref{fig:2dStacks}. The signal should be consistent with null for trial temperatures that match the true cluster temperatures, seen here for the 5-10 keV examples. For visualization purposes we only show measurements for five of the 38 trial temperatures. The data points are highly correlated, as shown in \cref{fig:cor_mat}.
    }
    \label{fig:data}
\end{figure}

In \cref{fig:2dStacks} we show 2D stacks of the ACT clusters at three trial temperatures using the combined ACT and \textit{Planck} data set, and in \cref{fig:data} we present the 1D azimuthally averaged profiles. The positive signal seen for $\bar{T}_e=0.1$\,keV indicates that the data favor a non-zero average cluster temperature, as demonstrated with simulations in \cref{fig:sehgal_sims}. Likewise, the large negative signal for $\bar{T}_e=20$\,keV shows that high temperatures are also disfavored. The ringing nature of the profile, e.g., the positive ring around the negative peak in the bottom panel of \cref{fig:2dStacks}, is driven by the high-pass filtering of the data and is not indicative of radial variations in the cluster temperatures. Future data will have the sensitivity to study radial variations. 

\begin{figure}
    \centering
\includegraphics[width=.48\textwidth]{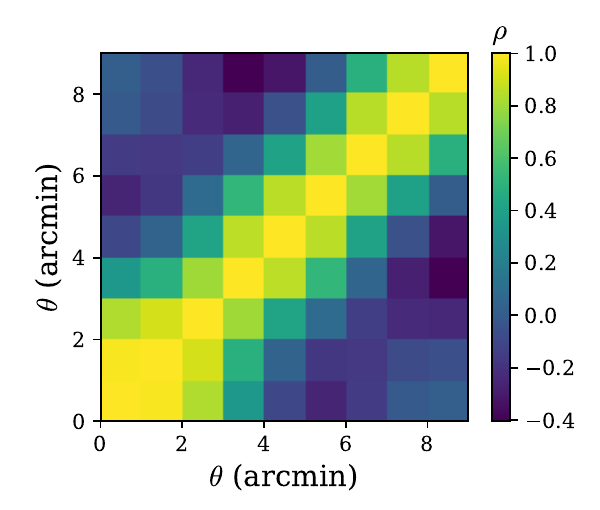}
    \caption{The correlation matrix for the measured 1D profiles of the g$^{(1)}$ map, \cref{eq:rsz_taylor}, at $0.1$\,keV stacked at location of the 4690 ACT clusters. The structure seen here is largely driven by radial binning, and the high- and low- pass filtering of the data. }
    \label{fig:cor_mat} 
\end{figure}

The correlation matrix of the $0.1\,$keV data points is shown in \cref{fig:cor_mat}. This is computed separately for each trial temperature using the method described in \cref{sec:pipelineOverview}. The weighting of the input frequency maps is different for each trial temperature as the frequency functions, \cref{eq:rSZ}, change with trial temperature. One result of this is that the variance of the data points is different for each trial temperature, as seen in \cref{fig:data}. However, we find that the structure of the correlation matrix is similar for each trial temperature. The structure of the correlation matrix is largely determined by the high- and low-pass filtering, and the stacking and binning operations.

\begin{figure}
    \centering
    \includegraphics[width=.49\textwidth]{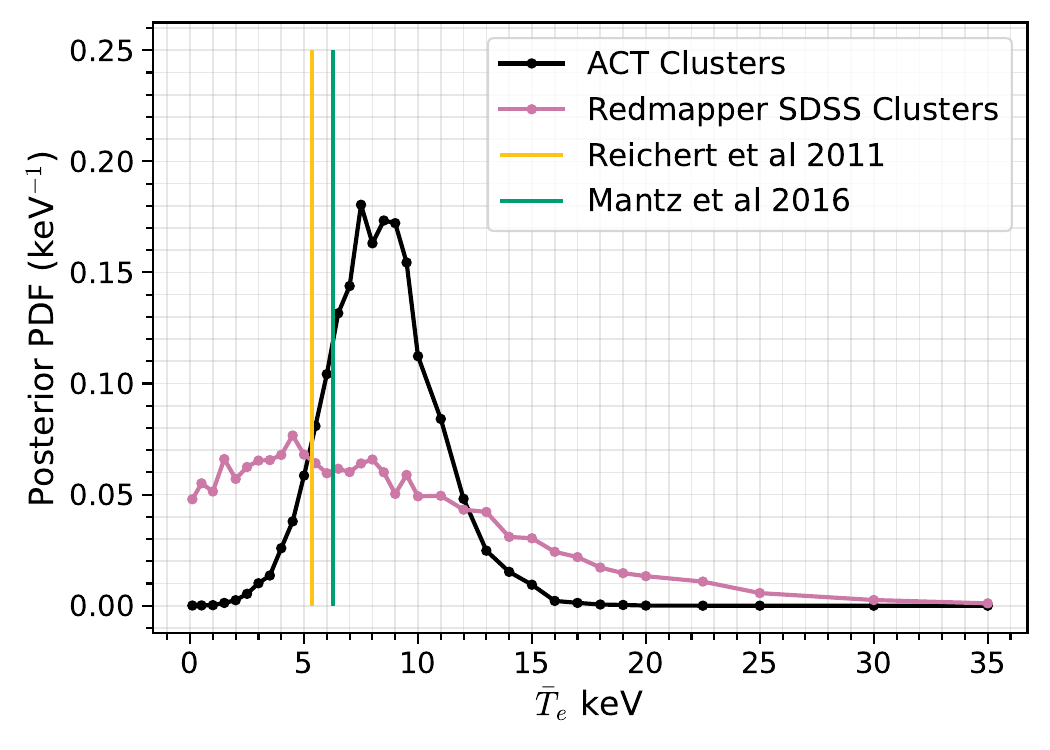}
    \caption{The temperature posterior for the average temperature of the ACT clusters (black). We plot the expected temperature computed using the $M_{500c}{-}T$ scaling relations from Chandra (orange) and XMM-Newton (teal)\citep{Pratt_2009,Reichert_2011,Mantz_2016}. The posterior is slightly non-Gaussian; it excludes zero at $>4.4$ Gaussian $\sigma$ and is consistent with both X-ray measurements. In pink, we show the posterior for a lower-mass optically selected cluster sample from SDSS (see \cref{sec:conclusions}). This result does not include the impact of instrumental systematic effects.}
    \label{fig:Tbar_posterior}
\end{figure}

Next we compute the posterior of the average temperature, \cref{eq:likelihood}. 
\cref{fig:Tbar_posterior} shows that the data measure a mean temperature of $8.5\,$keV $\pm 2.4$\,keV. From the non-Gaussian posterior we compute $p(T_e<0.1\mathrm{keV})<5.4\times10^{-6}$, equivalent to $4.4$ Gaussian $\sigma$. Alternatively we can compute the change in chi-squared between $\bar{T}_e=0.1$\,keV and the posterior mode. We find $\delta\chi^2=-16.2$, equivalent to $3.9\, \sigma$.


\subsection{Assessment of Potential Contaminants}\label{sec:contaminants}

To contextualize the importance of contaminants, consider the size of the rSZ signal relative to contaminant CIB and radio sources. A typical cluster in our catalog has an integrated flux of $\sim 50$\,mJy. A steep spectrum radio source with flux $10$\,mJy at 1.4\,GHz would contribute $0.5$ and $0.4$\,mJy at 90\,GHz and 150\,GHz respectively. A dusty point source with $15$\,mJy flux at $220$\,GHz would contribute  1.0 and 6.3\,\,mJy at 90\,GHz and 150\,GHz respectively. Below 217\,GHz the SZ effect is negative and these sources would infill the tSZ decrement. Given the relativistic tSZ signal is $\sim5\%$ of a cluster flux (i.e.\ $\sim2.5$\,mJy for a typical cluster), these sources could contribute non-trivial biases to the temperature measurements.  Larger electron temperatures lead to reductions of the tSZ decrement below the null frequency so infilling by contaminants, in the absences of mitigation, will bias measurements to higher temperatures. As described in \cref{sec:stacking}, the frequency dependence of the CIB is different from the rSZ effect and this is leveraged to mitigate the CIB contamination in the NILC. Clusters near bright radio sources are masked to mitigate radio contamination. This section explores the efficacy of these approaches.

As discussed in Ref. \citep{Dicker_2021,Dicker_2024}, many tSZ clusters can be significantly infilled by radio sources. For example, Ref. \citep{Dicker_2021} estimates that $\sim 13\%$ of clusters have their tSZ flux reduced by $>5\%$. More recently \citep{Dicker_2024} found that $\sim10\%$ of clusters have a radio source with flux $>1$mJy at 90\,GHz. Ref. \citep{Erler_2018,Orlowski_Scherer_2021} found a similar level of contamination from dusty star forming galaxies. This infilling is slightly larger than the typical size of the relativistic SZ corrections. Understanding how radio sources and the CIB contaminate our measurement depends on how the spectral shapes of these sources matches the rSZ signal, shown in Fig. \ref{fig:responses}. 
The state-of-the-art simulations, used in \cref{sec:sims} to validate that our method is not biased by these sources, are approximations of reality. Accurately modeling the CIB and radio sky signals is very challenging and so we perform a set of data-based tests to further validate our measurement.

\begin{figure}
    \centering
    \includegraphics[width=.49\textwidth]{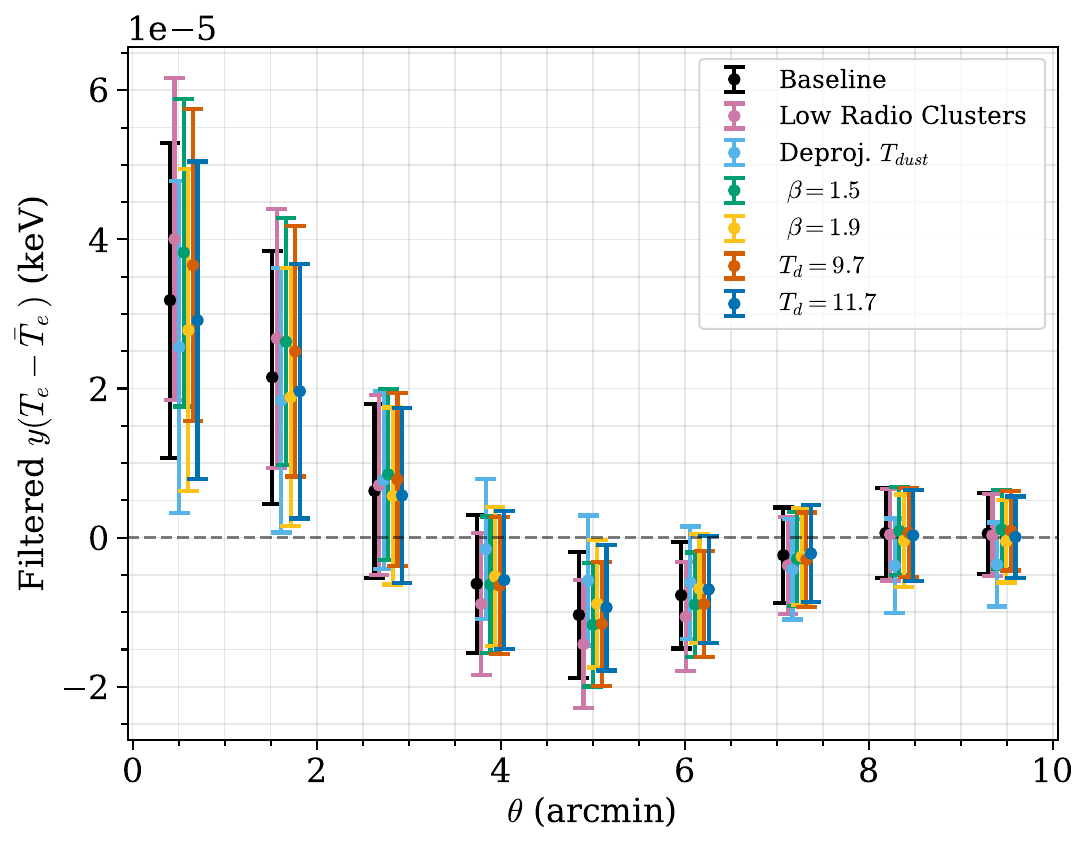}
    \caption{A set of data-based tests for the impact of contamination for the $\bar{T}_e=5$\,keV measurements. We verify that changes to the assumed CIB spectral parameters ($\beta^\mathrm{fiducial}=1.7$ and $T^\mathrm{fiducial}_d=10.7$\,K) lead to negligible shifts in our results (green, yellow, orange and dark blue). We also test that changing the CIB mitigation method to deproject the derivative of the SED with respect to dust temperature, instead of spectral index, leads to consistent results  (cyan). Finally, the pink points, show the measurement excluding clusters that have a radio source in RACS-mid catalog with flux within a 2.5$^\prime$ disc exceeding $10\,$mJy.  Note the effects of the filter at 5$^\prime.$  }
    \label{fig:ModellingTests}
\end{figure}

First, we consider the impact of different choices for modeling the CIB. Our procedure to remove the CIB (a constrained ILC that removes a modified blackbody, i.e. \cref{eq:greyBody}, and its derivative with respect to the spectral index) assumes an approximate form of the CIB spectrum. Whilst the moment expansion is designed to capture differences between the ``true" CIB spectrum and the approximate form, it is important to understand the sensitivity of our results to different modeling assumptions \citep[see, e.g., Refs.][for a detailed discussion]{Chluba_2017,Coulton_2023,McCarthy_2023b}. In this case we vary the CIB spectral index and dust temperature within a measure of their uncertainty from Ref. \citep{McCarthy_2023a}. In \cref{fig:ModellingTests} we report measurements at one trial temperature, $\bar{T}_e=5$\,keV; however, similar results are seen at other temperatures. We see the changes in the assumed parameters lead to changes that are much smaller than our measured errors. As a second test we instead deproject the derivative of the CIB spectrum with respect to the dust temperature, instead of the spectral index. As seen in \cref{fig:ModellingTests}, the results are also stable to this change. These results provide evidence that our measurement is robust to changes in our treatment of the CIB and is likely not strongly contaminated by the CIB.

Next we check for the impact of radio sources, such as active galactic nuclei (AGN) or radio galaxies \citep{de_Zotti_2010}.  To do this we identify clusters that have a total flux within a 2.5$^\prime$ disc that exceeds $10$\,mJy, as measured with the RACS-mid 1.4\,GHz data \citep{Duchesne_2023a,Duchesne_2023b}, and exclude these objects from our analysis. As seen in \cref{fig:ModellingTests}, this also leads to a negligible change in the measurement. Similar results are obtained if we use the NVSS \citep{Condon_1998}, VLASS \citep{Gordon_2021} or SUMSS \citep{Mauch_2003} radio data.

\begin{figure}
    \centering
    \includegraphics[width=0.48\textwidth]{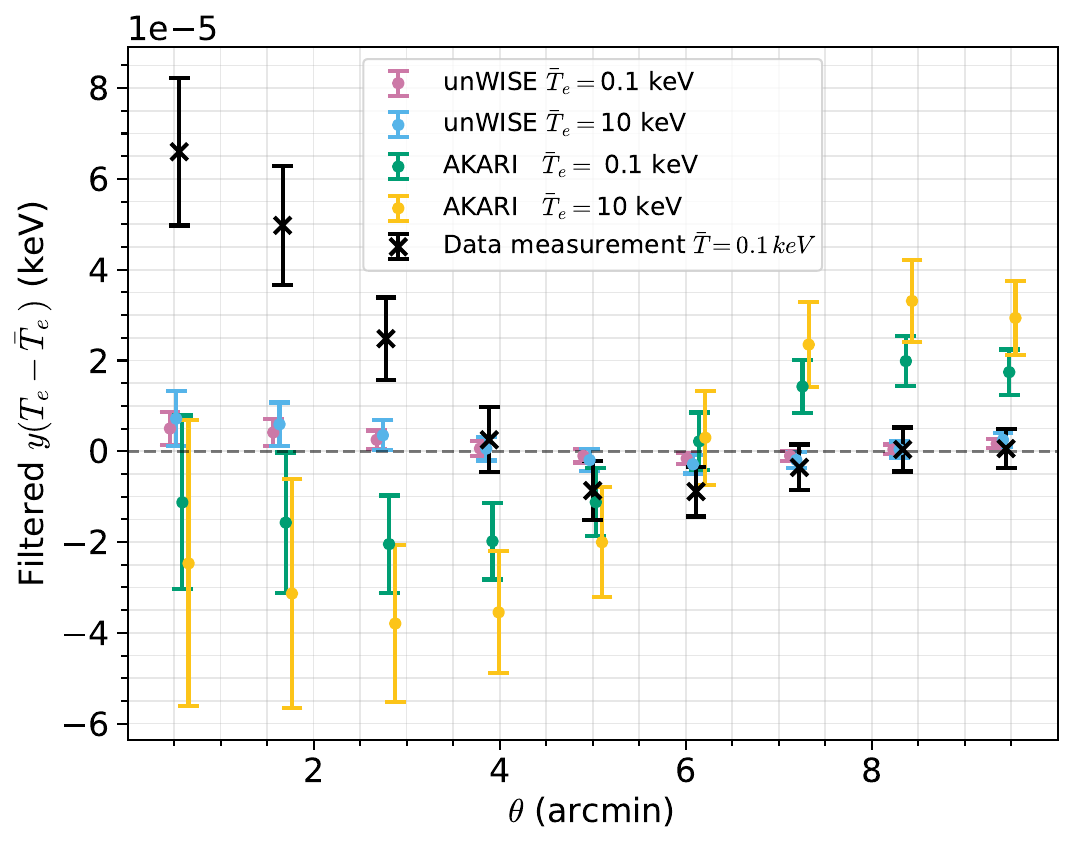}
    \caption{We perform two further data-based tests for the CIB contamination: we stack on {$\sim3800$} far-infrared objects detected by AKARI at $160\,\si{\micro\metre}$ \citep{Murakami_2007,Doi_2015} and $\sim 96,000$ galaxies from the \textit{unWISE} red catalogs \citep{Schlafly_2019,Krolewski_2020,Krolewski_2021}. If CIB sources were strongly biasing our cluster temperature measurements, when we stack on these CIB tracers we would expect to see a large signal even though we are far from any massive clusters. Measurements of both the unWISE and AKARI sources are consistent with zero in the central region providing further evidence that there are no large CIB biases. There is evidence of a bias at $3^\prime$ from AKARI; however the differences in the AKARI sample to the cluster sample  and difference between the shape of this bias and the signal mean that it is unlikely to bias our temperature measurement, see \cref{sec:contaminants} for further discussion.  To contextualize these results,  we show our measurements at the locations of SZ clusters, with $\bar{T}_e=0.1$\,keV, in black. }
    \label{fig:akari_sources}
\end{figure}
\begin{figure}
    \centering
    \includegraphics[width=0.48\textwidth]{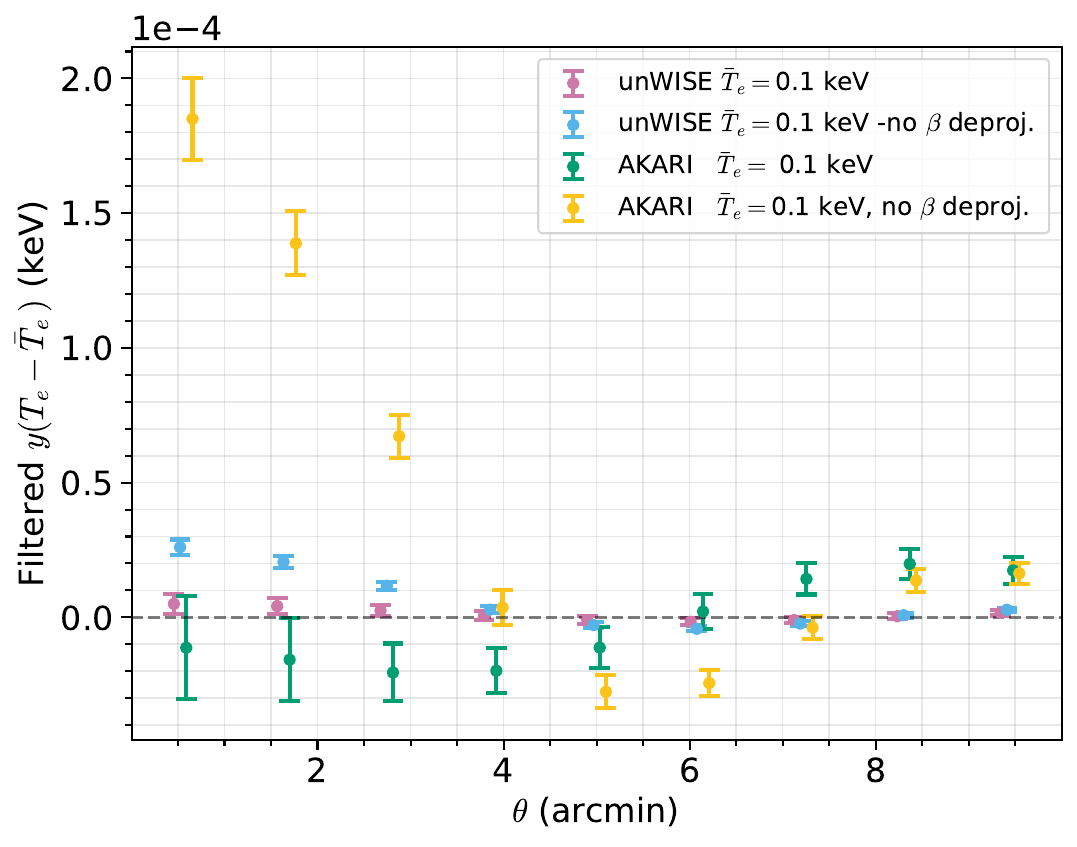}
    \caption{A test of the CIB mitigation methods using the same AKARAI and unWISE galaxy samples as \cref{fig:akari_sources}. Whilst neither catalog ideally represents contaminant CIB sources, stacking on these sources is a non-trivial test. Here we compare our baseline method to the case when we remove the second CIB mitigation method, the derivative deprojection. In the latter case, we  see a significant bias (blue and yellow). This is both a test demonstrating the need for our full CIB mitigation method and a validation that the two samples are tracers of the CIB.}
    \label{fig:akari_sources_nodBeta}
\end{figure}
\begin{figure}
    \centering
    \includegraphics[width=0.48\textwidth]{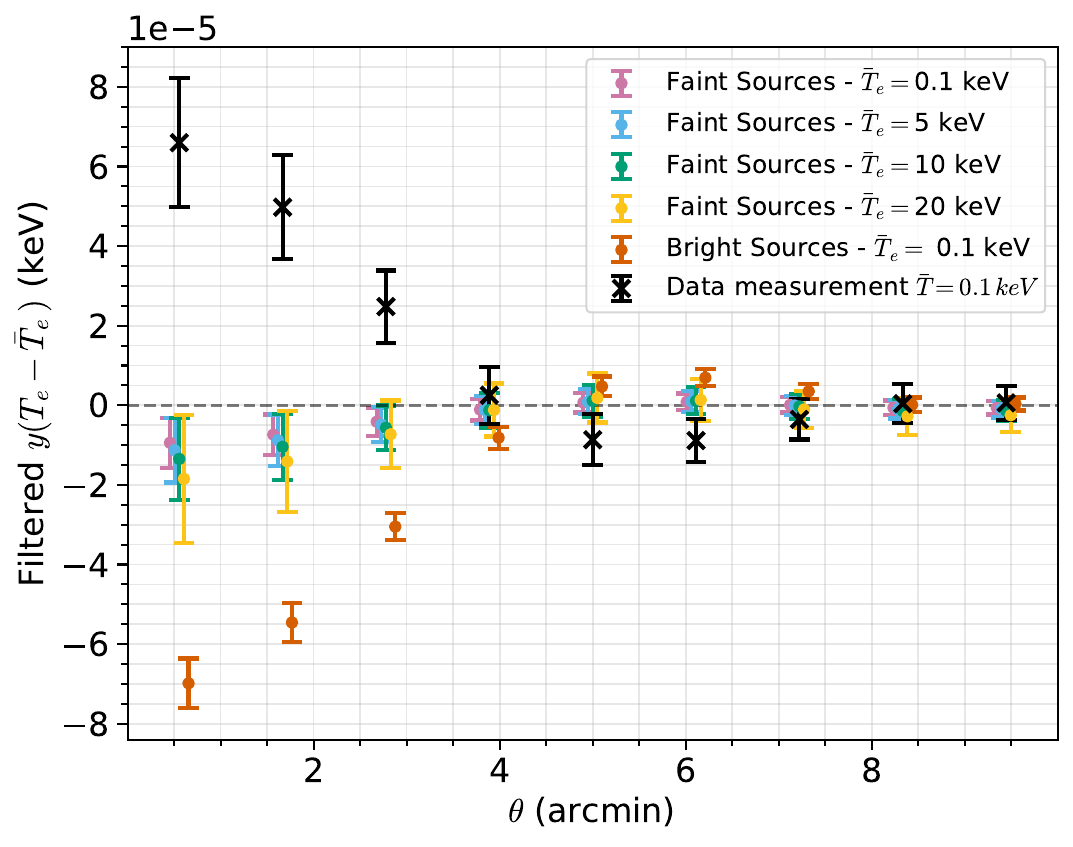}
    \caption{A test for any residual bias from radio galaxies. We apply our analysis method to a stacks of objects from the Rapid ASKAP Continuum Survey 1367.5\,MHz radio catalogs \citep{Duchesne_2023a,Duchesne_2023b}. The faint sample contains $\sim 31,000$ objects with a mean flux within a $2.5^\prime$ disc of $11$\,mJy. No significant bias is seen for this sample. Using the same ASKAP catalog, we find that the mean flux of radio objects in a $2.5^\prime$ disc of the ACT clusters is $6.9$\,mJy. In orange, we repeat this analysis for a sample of $33,000$ brighter radio objects, with mean disc flux of $35$\,mJy. A large bias is seen for this sample demonstrating the potential bias from radio contamination. Clusters with such bright objects are excluded from our analysis to avoid any bias, but this test allows us to understand the structure of biases from unmitigated radio galaxies. Note that, as discussed in the text, the spectral indices of the bright sample are different than the faint sample. As in \cref{fig:akari_sources}, we show our measurements at the locations of SZ clusters, with $\bar{T}_e=0.1$\,keV, in black. }
    \label{fig:askap_sources}
\end{figure}

To further test the impact of the CIB and radio galaxies we perform two stricter tests. The idea of these tests is to perform our analysis replacing the cluster catalog with a catalog of either dusty star-forming galaxies or radio sources. These measurements provide a direct data-based assessment of any potential biases.

There are currently no catalogs of CIB sources that cover the full ACT footprint with the required sensitivity. Thus, we use two proxy catalogs to assess the bias: the first is the catalog of bright far infra-red sources from the AKARI satellite \citep{Murakami_2007,Doi_2015}. The AKARI satellite observed the full-sky sky between 1.7 and $180\,\si{\micro\metre}$ and has detected the CIB~\citep{Matsuura_2011}. Unfortunately, the far-infrared catalog is not ideally suited for our purpose as the sources are primarily at low redshift, $z\lessapprox 0.1$, and are contaminated with Galactic sources~\citep[e.g.,][]{Pollo_2012}.\footnote{We repeated our stacking using the IRAS catalogs \citep{IRAS_1998} and find similar results to AKARI. } We use $\sim3,800$ AKARI sources that have fluxes $<30$\,Jy and are at least $6^\prime$ from detected SZ clusters, $10^\prime$ from bright radio galaxies (in the ASKAP catalogs) and $2^\prime$ from subtracted point sources.  The second catalog is the unWISE red catalog from Refs. \citep{Lang_2014,Schlafly_2019,Krolewski_2020,Krolewski_2021}. The unWISE catalogs are created from the extended observations of the Wide-field Infrared Survey Explorer~\citep{Wright_2010,Meisner_2018} and the red galaxy sample,\footnote{Similar results are obtained for the green and blue samples.} which we use, comprises a dense selection of high redshift galaxies. Due to the similar galaxy masses and redshifts, many of these unWISE galaxies likely contribute to the CIB and thus are a useful catalog for testing CIB biases \citep{planck2013-pip56,Maniyar_2021,Kusiak_2022,Kusiak_2023,Yan_2023}. We use $\sim 96,000$ objects that are randomly selected from the catalog and we apply the same cuts as with the AKARI data to avoid masked regions, known clusters and point sources.\footnote{An alternative approach would have been to use bright sources detected in the \textit{Planck} high-frequency maps; however, we are unable to do this as we subtract all detected sources in our component-separation pipeline (as described in Ref. \citep{Coulton_2023}) so remove any sources detected by \textit{Planck}.} 

The results of performing our temperature analysis at the location of these objects is shown in \cref{fig:akari_sources}. Measurements of both the unWISE and AKARI stacks near the center are statistically consistent with zero so no significant biases are seen.  We tested different flux cuts on the objects in the AKARI catalog and consistently found no central bias. For the AKARI stacks a small ($\sim2\,\sigma$) bias can be seen at  $>3^\prime$ from the center. This apparent bias at $\sim3^\prime$ could be a statistical fluctuation or be a sign of extended galactic objects in the AKARI catalog. The shape is very different from our expected signal, see \cref{fig:sims_rSZ}, and so could be easily identified in the measurement. We see no evidence of such a signal in our measurement and the vanishing central bias implies that it is not from the CIB. Thus, we conclude that this is likely a feature of the AKARI sample, which is very different to the cluster sample. Given the caveats with the two catalogs this could be a trivial result. To test this we relax the CIB mitigation methods, by removing the deprojection of the derivative of the modified blackbody. In this case, shown in \cref{fig:akari_sources_nodBeta}, we see a large bias. This implies 1) that sources within the catalog do represent CIB-like objects that could bias our measurement and 2) that our mitigation method removes these. The AKARI sources are brighter than those expected in the clusters and so they provide a powerful test of potential biases.

To test for radio contamination, we repeat this process using objects within the RACS-mid catalog. We select $\sim 31,000$ sources with flux $<10$\,mJy and only include objects away from known clusters, subtracted point sources and inpainted regions (in an identical manner to the AKARI sources). The ASKAP angular resolution ($\sim$ arcseconds) is significantly higher than ACT, so many point sources fall within the effective area of our analysis. Thus, to compute the bias to our measurements we consider the total flux within a disc of radius $2.5^\prime$. For this sample of RACS-mid objects the total flux within a 2.5$^\prime$ disc is 11\,mJy, which is slightly larger than the average RACS-mid flux within 2.5$^\prime$ of the clusters in our sample, 6.9\,mJy.\footnote{Note we do not expect the mean fluxes to match as some tSZ clusters have no detected radio sources in RACS.} \cref{fig:askap_sources} shows slight evidence of a small, negative bias from this sample of radio sources. As this bias is smaller than the size of the error bar for our measured signal (shown in black), and would act to increase the SNR of our measurement, we do not explore further mitigation methods.  We repeat this analysis with a brighter sample of radio sources, which have a mean flux within a 2.5$^\prime$ disc of $35$\,mJy. The results for this sample, also shown in \cref{fig:askap_sources}, demonstrate that unmitigated contamination from radio galaxies would significantly bias our inference. Note that there is a subtle selection effect at play that means that the physical properties of these two RACS-mid samples are different. To understand this consider a 35\,mJy flat-spectrum source in the bright sample. This source would be above the ACT point source detection thresholds in the f090 and f150 maps. Therefore this source would be subtracted and not present in our analysis of RACS-mid sources, which avoids subtracted sources. A steep spectrum source on the other hand would likely not be detected and so would be present. For the faint samples, both flat-spectrum and steep-spectrum sources would be below the ACT thresholds and so present in the stacks.

In summary, we developed several approaches to assess potential biases from cosmic infrared background and radio galaxies. Taken together they provide strong evidence that our measurement is not significantly biased by either contaminant.

\begin{figure}
    \centering
    \includegraphics[width=0.5\textwidth]{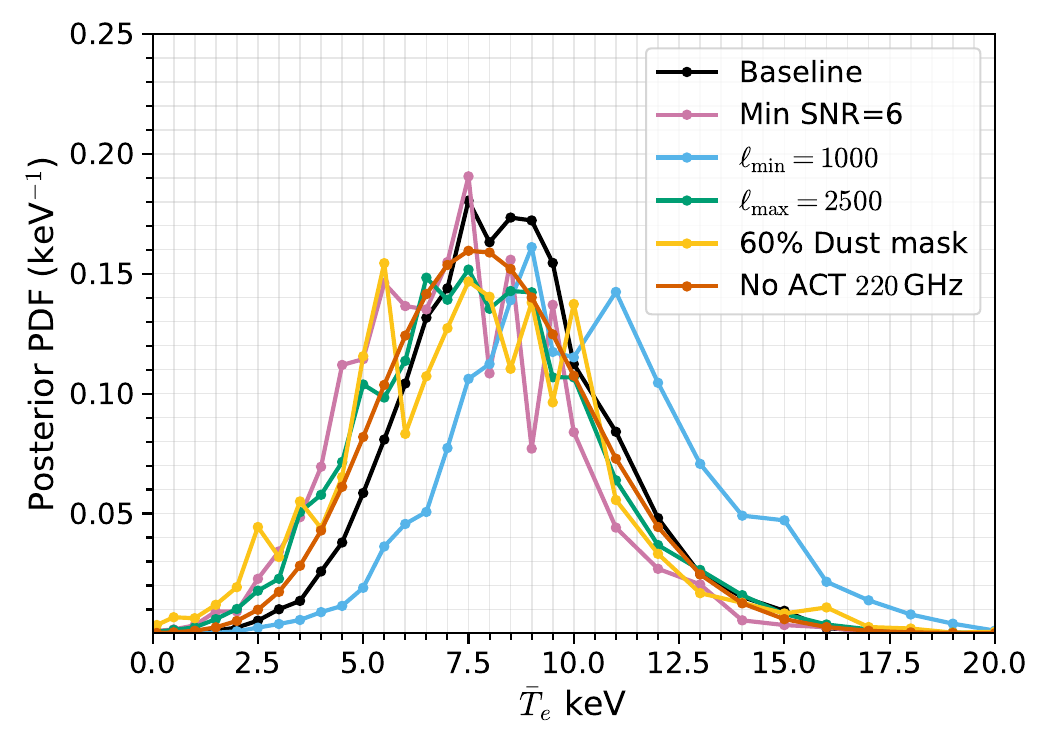}
    \caption{The impact of different analysis choices on the inferred temperature posterior. In pink we show the impact of only using the 2693 ACT clusters with SNR$>6$ (cf. SNR$>5$ for the fiducial). In blue we show the posterior for an increased minimum scale $\ell_\mathrm{min}=1000$ (cf $\ell_\mathrm{min}=500$ for the fiducial) while in green we show the result after reducing the maximum scales used $\ell_\mathrm{max}=2500$ (cf. $\ell_\mathrm{max}=3200$ for the fiducial). In yellow, we show the measurement when masking regions of the sky that may be strongly contaminated with dust emission. The mask removes regions of the sky brighter than the 60$\%$ quantile of the \textit{Planck} 857\,GHz map, which we use as a dust proxy. There are 2918 clusters within the reduced mask. Finally, in orange we show the result of excluding the ACT f220 data, the channel that is hardest to calibrate. For reference, in black we show our baseline analysis. Each of these cuts removes significant amounts of data and all posteriors exclude $\bar{T}_e=0.1$\,keV with probability $>0.997$. This plot assumes no instrumental systematic errors. }
    \label{fig:differentCuts}
\end{figure}

\subsection{Data Consistency Tests}\label{sec:dataConsistency}
Next we assess the impact of our analysis choices. \cref{fig:differentCuts} shows how our temperature constraint changes when we change the SNR cut applied to the catalog, the minimum and maximum scales used in the analysis, the area of sky used and if we exclude f220 data. The tests of the scales and SNR cut are useful consistency checks of our measurements and provide some measure of the scales and types of object driving our constraining power. The sky area test is designed to test for contamination from Galactic foregrounds. We use the \textit{Planck} 857\,GHz map, which is otherwise excluded from our analysis due to its large calibration error, to construct a mask of regions with bright dust emission. The consistency between this measurement and our baseline result suggests that we are not strongly contaminated by Galactic emissions.  Similarly we find that either changing the scales used in the analysis or increasing the SNR cut leads to consistent results. Finally, we explore the impact of removing the ACT f220 data. The ACT f220 data are hardest to characterise and most difficult to calibrate so most at risk of systematic effects. We see that removing these data does not significantly impact our measurement. Interestingly, the resulting posterior is much smoother than the other null tests. This may arises as f220 is both very noisy and an important channel, as the tSZ null location (around 217\,GHz) is sensitive to relativistic corrections. All these tests find $\bar{T}_e>0.1$\,keV with probability $>0.997$ and so provide further evidence for the robustness of our measurement. Each one of these tests reduces the amount of data, and thus the constraining power. While including more objects or smaller scales will help increase the SNR of the measurements, the measurement is already limited by instrumental systematic errors, which we discuss next, rather than statistical errors.

\begin{figure}
    \centering
    \includegraphics[width=.49\textwidth]{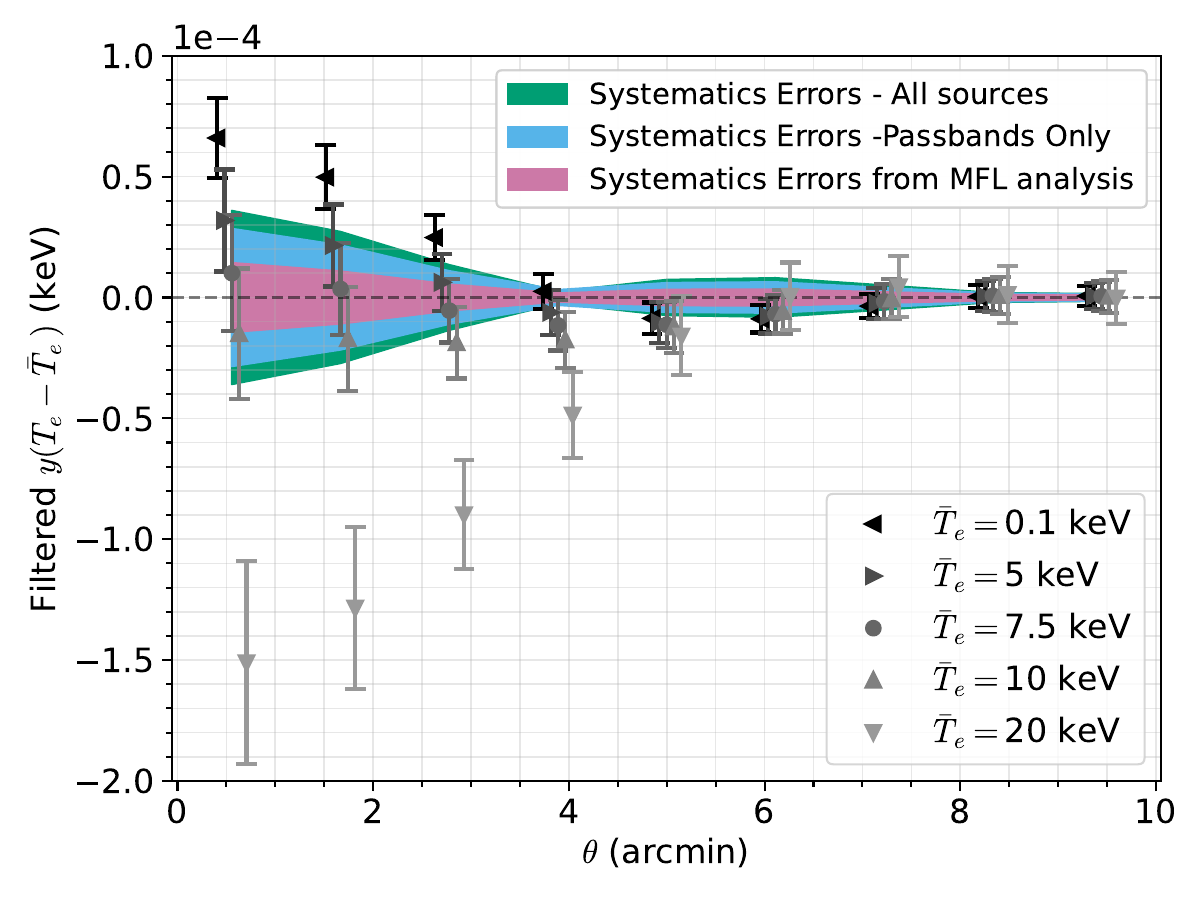}
    \caption{A comparison of the instrumental systematic errors in our measurement to the statistical error. All the error bars and bands denote $1\sigma$ (68\%) regions. The green band shows the total systematic error, whilst the blue band shows the systematic error from the passband uncertainties only. The passband uncertainties dominate the systematic error budget. The data 1D profiles, measured with five trial temperatures, with statistical errors are shown in black. Direct measurements, such as Fourier Transform Spectrometer passband measurements, are not the only way to constrain the instrument's properties and other paths could reduce the impact of these systematic effects. As one example, we show in purple the rSZ systematic errors when we use constraints on the passbands from a forecast multi-frequency likelihood (MFL) CMB power spectrum analysis of ACT DR6 data. Using the MFL analysis would significantly mitigate the impact of systematic effects on rSZ studies.  All error bands were computed at $\bar{T}_e=5\,$keV and are plotted around zero to aid comparison. }
    \label{fig:systematics}
\end{figure}

Finally we assess the impact of instrumental systematic effects. Following the procedure described in Ref.~\citep{Coulton_2023}, we sample uncertainties in the beams, passbands and calibrations and repeat our analysis. The distribution in the resulting measurements gives a measure of the impact of these effects and an estimate of the systematic error in our analysis. The result, shown in \cref{fig:systematics}, shows that systematic errors are larger than the statistical uncertainties. If we repeat this exercise and only sample passbands we find that these dominate the systematic errors. The passband uncertainties on \textit{Planck} maps, as reported in Refs \citep{planck2013-p03d,Planck_int_LVII}, are negligible and so this result is driven by the ACT uncertainties. ACT passband uncertainties have two components: a shape uncertainty and a central frequency uncertainty. The error in the central frequency is the most important. The shifts used here are Gaussian shifts around the mean with standard deviations of $\sim 1.4$\,GHz at f090, $\sim2.2$\,GHz at f150 and $\sim 3.5$\,GHz at f220. The uncertainty in the f150 channel carries the most weight for our experimental configuration. The systematic errors used here are from direct measurements of instrument properties (e.g., Fourier Transform Spectrometer measurements of the passbands); however, instrumental parameters can be constrained from CMB observations themselves and could provide improvements upon existing measurements. As an example, passbands and calibration parameters are constrained in CMB multifrequency likelihood (MFL) analyses \citep[e.g.,][]{Choi_2020}. In \cref{fig:systematics}, we show how the systematic errors change if we instead use ACT DR6 forecast constraints on systematic parameters.\footnote{MFL typically do not include the rSZ effect and it needs to be investigated whether this would impact this analysis.} The systematic errors on the rSZ measurements are reduced by a factor of $\sim 2$. This provides evidence that with data-based calibrations, systematic errors can be sufficiently controlled for rSZ studies. This analysis was performed only for a single temperature, due to the computational expense; however, we expect similar results at other temperatures.  We discuss the implications of this result on future measurements in \cref{sec:conclusions}.

\section{Constraints on scaling relations}\label{sec:temp_evolution}

\begin{figure}
    \centering
    \includegraphics[width=.49\textwidth]{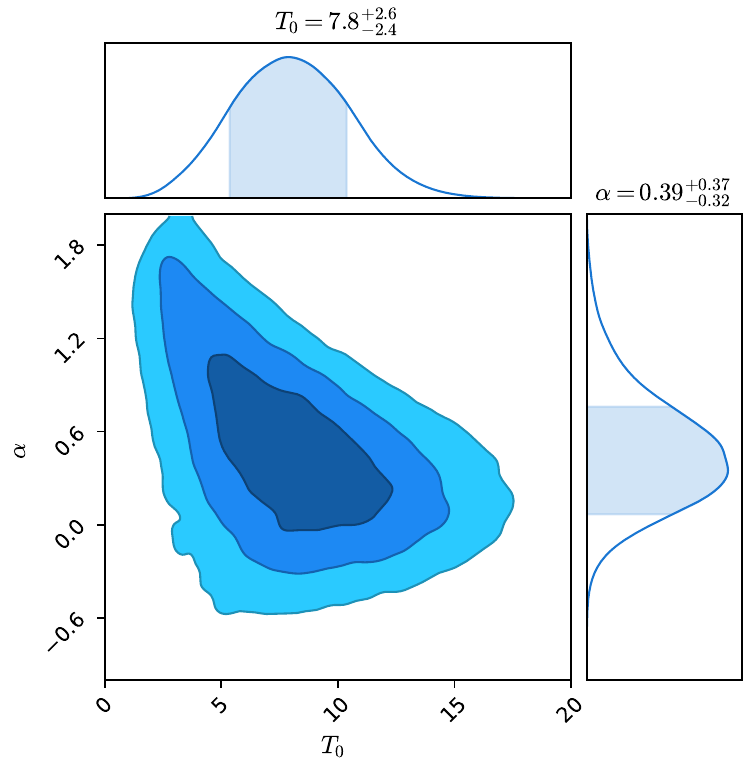}
    \caption{The posterior for the pivot temperature ($T_0$) and mass power law index, $\alpha$ relating the temperature to the observed integrated Compton-$y$ signal, $y_{500}$: $T_e=T_0 (y_{500}/y^0_{500})^\alpha$ with a pivot of $y_{500}^0=1\times 10^{-4}$. The contours denote the 1, 2 and 3 $\sigma$ confidence intervals. As expected, this analysis favors positive values of $\alpha$ which means clusters with larger $y_{500}$ are hotter. 
    }
    \label{fig:scaling_relations}
\end{figure}

Future rSZ measurements will allow inferences on how temperatures evolve with mass and redshift. To demonstrate this, we consider how well this sample of clusters could be used to constrain temperature-Compton-$y$ scaling relations. We divide the cluster sample into 5 bins, each with $>500$ galaxy clusters, based on the value of the integrated Compton-$y$ signal within $r_{500}$, $y_{500}$. For each bin we compute the temperature posterior. We then fit the dependence of the cluster temperature on the integrated Compton-$y$ value. We parameterize this relationship as
\begin{align}
    T_e = T_0 \left(\frac{y_{500}}{y_{500}^0}\right)^\alpha.
\end{align}
where $T_0$ is the pivot temperature, $y^0_{500}=1\times 10^{-4}$ is the pivot value of $y_{500}$, and $\alpha$ is the power law index.
Explicitly we have
\begin{align}
    P(T_0,\alpha|d) \propto \prod\limits_{\mathrm{Sample\,i}} P(d_i|\bar{T}_i) P(\bar{T}_i| T_0,\alpha) P(T_0)P(\alpha)
\end{align}
where $P(\bar{T}_i| T_0,\alpha)$ is a deterministic relationship computed by evaluating the mass-$y$ relationship for each cluster within the sample. We assume a flat prior of $-4<\alpha<4$ and $0\,$keV$<T_0<35$\,keV. Given the small parameter space we simply evaluate the posterior on a 2D grid using 400 points per dimension. This analysis does not account for the systematic errors, discussed in \cref{sec:dataConsistency}. 
The resulting constraints are shown in \cref{fig:scaling_relations}. Given the limited signal-to-noise of our measurement, this is primarily demonstrative of the future utility of such measurements. However, note that this analysis also provides strong evidence for a non-zero cluster temperature as $p(T_0<0.1\,\mathrm{keV})=3\times10^{-6}$ or $>4.5$ Gaussian equivalent $\sigma$, and evidence that clusters with larger values of $y_{500}$ are hotter.  Both of these match expectations; for example, self similar models predict $\alpha=2/5$ and simulations find $0.31\lesssim\alpha\lesssim 0.38$ \citep{Lee_2020,Lee_2022}.

\section{Discussions and outlook}\label{sec:conclusions}
In this work we used relativistic corrections to the tSZ effect to measure the average temperature of 4690 Compton-$y$ selected clusters. We find the mean temperature to be $8.5\pm 2.4$\,keV, with a comparable systematic error. This measurements is consistent, but slightly higher, than past rSZ measurements. For example, Ref. \citep{Erler_2018} used \textit{Planck} data to find $\langle T_e\rangle=4.4^{+2.1}_{-2.0}$ for the full sample of \textit{Planck} clusters and $\langle T_e\rangle=6.0^{+3.8}_{-2.9}$ for the hottest 100 clusters. Similarly Ref. \citep{Remazeilles_2024} found $\langle T_e\rangle=4.9^{+2.6}_{-2.6}$ when they analyzed \textit{Planck} maps and the full \textit{Planck} cluster sample. Compared to Ref. \citep{Remazeilles_2024}, we use a larger cluster sample, 4960 vs 795 clusters, with a higher mean redshift, 0.59 vs 0.23, and slightly lower median mass, $3.6\times 10^{14}M_\odot$ vs $4.75\times 10^{14}M_\odot$. The differences could be indicative of redshift evolution in the temperature\citep{Churazov_2015} or, more likely, a result of statistical fluctuations or systematics. Given that we use the same method as Ref. \citep{Remazeilles_2024} and add in high resolution data, one may have expected a more dramatic improvement in the measurement. However, the CIB mitigation methods that we add in this work to ensure robustness dramatically reduce the constraining power. An interesting question for future work is whether alternative analysis methods, such as measuring the spectrum of individual (or stacks of) clusters , could provide more precise constraints.

We performed a suite of tests to assess the impact of foreground contamination, finding no significant biases from the CIB or Galactic sources. These tests show that the constrained ILC with the moment expansion are effective at mitigating CIB contamination - at the cost of increased noise. When stacking on radio sources we find that they could bias our measurement at a $\lesssim 0.5\,\sigma$ level. Accurately assessing the bias from radio galaxies is difficult as the level of radio contamination present in our cluster sample is uncertain \citep{Dicker_2021}. Further, the dominant radio sources, flat-spectrum radio quasars and blazars, vary significantly over time \citep{Trippe_2011,Chen_2013,Richards_2014}. Fortunately, future surveys will be less impacted by radio sources as they will have broader frequency coverage than the ACT data used here. The resulting low frequency maps will allow a better separation of radio sources from the SZ effects within the NILC. 
Alternatively, including external data sets could be a powerful way of ensuring there is no radio contamination. One approach would be to include radio observations in the NILC to directly remove radio contamination. A second approach would be to use high resolution millimeter observations to directly quantify the radio contamination at the observed frequencies. In this later direction, Ref. \citep{Dicker_2024} already used MUSTANG2 observations to show that, for their sample of 243 clusters, only $\sim 10\%$ of clusters have sources with flux $>1$\,mJy at 90\,GHz. This already suggests that radio contamination may not be large; however a larger, more systematic study is needed to fully understand radio contamination in rSZ measurements.

Finally, we investigated the impact of instrumental systematic effects and found that the resulting systematic error is already limiting this analysis.The key instrumental systematic effect is the uncertainty in the central frequency of the passbands. This is a potential concern for future surveys as this will not average down with more objects or increased depth. For example, whilst upcoming experiments will have many more detector arrays than ACT, the dominant passband systematic error will not necessarily average down when combining data from multiple arrays. This motivates a range of design choices that allow for better understood instrumental effects, as well as the exploration of novel means of measuring the beams, passbands and map calibration. As one example of this, we showed that with ACT CMB power spectrum constraints on the passbands the impact of systematics errors would be significantly mitigated.

It would be interesting to compare our temperature measurements to the current state-of-the-art method: X-ray observations. Unfortunately public X-ray temperatures are currently available for only a small subset of the objects used here. Instead, we can use X-ray derived mass-temperature relations to infer the expected, $y$-weighted temperature of our clusters. First we convert the observed cluster Compton-$y$ to mass, as described in Ref. \citep{Hilton_2021}, then we convert these to temperatures using the $M_{500c}{-}T$ relations from Ref. \citep{Mantz_2016}, computed with Chandra data, and from Refs. \citep{Pratt_2009,Reichert_2011}, obtained from XMM-Newton data. These are weighted in the stack by their measured central Compton-$y$ value, $y_c$. Note that the first step of this analysis, converting from Compton-$y$ to M$_{500c}$, requires us to use only confirmed clusters with redshifts and requires an assumed $y_c$--M$_{500c}$ relation, for which we use the method from \citep{Sifon_2016,Hilton_2018}. As seen in \cref{fig:Tbar_posterior}, both these X-ray derived mean temperatures are consistent with, but slightly lower than, our measurements.

This work highlights the power of small-scale CMB observations to study galaxy clusters' temperatures. Upcoming  experiments and facilities, such as SO, CMB-S4 \citep{S4_2016}, CCAT \citep{Parshley2018,CCAT-Prime_2023}, CMB-HD \citep{2022arXiv220305728T}, and the Atacama Large Aperture Submillimeter Telescope \citep[AtLAST][]{DiMascolo2024, Mroczkowski_2024}, will offer significant advancements through their higher sensitivity and extended frequency coverage. Both aspects are important: higher sensitivity will increase the precision, whilst extended frequency coverage helps mitigate biases and boosts the signal as the rSZ signal is stronger at higher frequencies. These results can complement X-ray observations by extending cluster temperature measurements to higher redshift. As an example of what could be learnt, we constrained the Compton-$y$-temperature scaling relation. With high precision measurements, this will provide informative measurements of how temperatures evolve with mass and redshift \citep{Churazov_2015}. This approach can also be used to characterize non-SZ selected cluster samples. As a demonstration of this we constrain the mean temperature of the 9675 SDSS DR8 redMaPPer clusters~\citep{Rykoff_2014} that fall within the ACT footprint. The resulting temperature constraint is shown in \cref{fig:Tbar_posterior}. The average mass of these objects is lower than the ACT sample and, as our signal scales with Compton-$y$, the temperature constraint is significantly weaker. However, this will also dramatically improve for future surveys. Together, upcoming measurements will allow detailed studies of cluster thermodynamics and evolution.  

\acknowledgments
We are very grateful to Jens Chluba, Mathieu Remazeilles and Anthony Challinor for insightful discussions. We also thank 
Martine Lokken for detailed comments on the manuscript. ADH acknowledges support from the Sutton Family Chair in Science, Christianity and Cultures, from the Faculty of Arts and Science, University of Toronto, and from the Natural Sciences and Engineering Research Council of Canada (NSERC) [RGPIN-2023-05014, DGECR-2023- 00180]. CS acknowledges support from the Agencia Nacional de Investigaci\'on y Desarrollo (ANID) through Basal project FB210003. The Flatiron Institute is supported by the Simons Foundation.

Support for ACT was through the U.S.~National Science Foundation through awards AST-0408698, AST-0965625, and AST-1440226 for the ACT project, as well as awards PHY-0355328, PHY-0855887 and PHY-1214379. Funding was also provided by Princeton University, the University of Pennsylvania, and a Canada Foundation for Innovation (CFI) award to UBC. ACT operated in the Parque Astron\'omico Atacama in northern Chile under the auspices of the Agencia Nacional de Investigaci\'on y Desarrollo (ANID). The development of multichroic detectors and lenses was supported by NASA grants NNX13AE56G and NNX14AB58G. Detector research at NIST was supported by the NIST Innovations in Measurement Science program. Computing for ACT was performed using the Princeton Research Computing resources at Princeton University, the National Energy Research Scientific Computing Center (NERSC), and the Niagara supercomputer at the SciNet HPC Consortium. SciNet is funded by the CFI under the auspices of Compute Canada, the Government of Ontario, the Ontario Research Fund–Research Excellence, and the University of Toronto. We thank the Republic of Chile for hosting ACT in the northern Atacama, and the local indigenous Licanantay communities whom we follow in observing and learning from the night sky.

\appendix
\section{Constructing an uninformative prior}\label{app:prior_choice}
In the main text we state that we use an uninformative prior such that the posterior is 
\begin{align}
    \ln P(\bar{T}_e|d) \propto -\mathbf{d}C^{-1}\mathbf{d}.
\end{align} 
Given our Gaussian likelihood this choice corresponds to a prior of $p(\bar{T}_e)=\ln|C|$, where $|C|$ is the determinant of the covariance matrix. In this appendix we explore why we choose this prior over a uniform prior and why we consider this to be ``uninformative".

The approximate log likelihood is given as 
\begin{align}
   2 \ln\mathcal{L}(d|\bar{T}_e)=- \mathbf{d}C^{-1}\mathbf{d} - \ln|C|.
\end{align}
Importantly, the covariance matrix depends on the trial temperature. This dependence arises not from the signal, as the measurements are dominated by instrumental noise, but rather because some temperatures are easier to measure and hence have lower noise. This occurs as different temperatures are constrained with different weights of the input frequency channels, and the different frequencies have different noise levels. Thus, the temperature dependence of the covariance matrix is set by the instrument. A consequence of this term is that temperatures with lower measurement noise are favoured by the likelihood, irrelevant of the true temperature. For our configuration, lower temperatures are better measured. We introduce our ``uninformative prior" as $p(\bar{T}_e) = \ln|C|$, which removes this dependence. This is desirable as the lower measurement noise is set by the instrument, not the cluster signal, and thus tells us nothing about the ``true" cluster temperature.

To demonstrate these effects and the value of our ``uninformative" prior, we perform a set of mock analyses. We generate a set of mock data such that 
\begin{align}
    d(\mathbf{n})= y(\mathbf{n})\left[T_\mathrm{true}-\bar{T}_e \right] +n(\mathbf{n}),
\end{align}
where $n$ is noise drawn from the covariance matrix used to analyze our data, $C(\bar{T}_e)$. We choose a Gaussian model for $y(\mathbf{n})$ (i.e. $y(\mathbf{n})=y_0/\sqrt{2\pi\sigma^2}\exp{\left[-0.5 |n|^2/\sigma^2\right]}$), with width $3^\prime$. We consider two amplitudes for the Gaussian: 1) a large signal model that has a large normalization of the Gaussian (integrated $y3\times 10^{-5}$) and 2) a very small signal model, with a small normalization (integrated $y3\times 10^{-7}$).  We analyze these with a flat temperature prior and with our uninformative prior. For each scenario we generate 1000 mock data vectors and analyze these. 

\begin{figure*}
    \centering
  \begin{subfloat}[Flat prior]{\includegraphics[width=.47\textwidth]{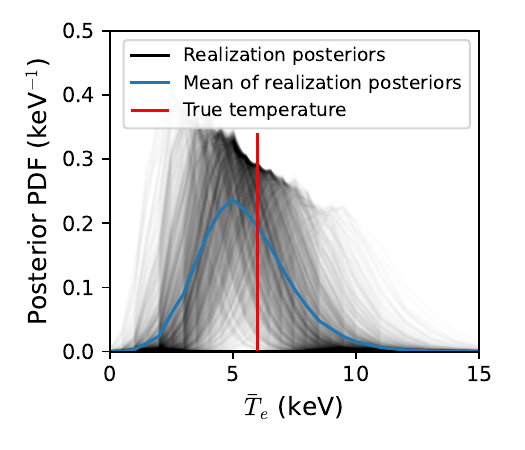} }
  \end{subfloat}
  \hfill
  \begin{subfloat}[Uninformative prior]{\includegraphics[width=.47\textwidth]{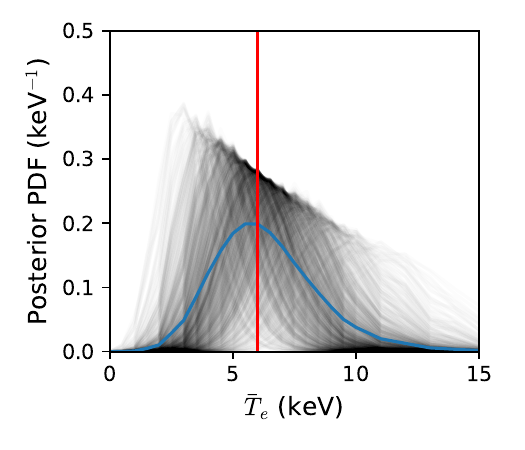} }
  \end{subfloat}
    \caption{The posteriors for 1000 mock analyses of a bright cluster sample  data vector. The posterior for each realization is shown in black and the mean of the posteriors is shown in blue. The data vector has a true temperature of $T_e=6\,$keV, denoted by the red line. We see that the mean posterior for the flat prior (left) peaks below the input signal, whereas the uninformative prior (right) recovers the true signal. }
    \label{fig:example_largeAmp} 
\end{figure*}

\begin{figure*}
    \centering
  \begin{subfloat}[Flat prior]{\includegraphics[width=.47\textwidth]{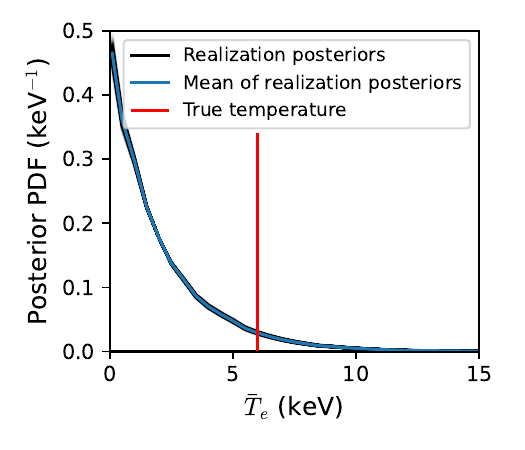} }
  \end{subfloat}
  \hfill
  \begin{subfloat}[Uninformative prior]{\includegraphics[width=.47\textwidth]{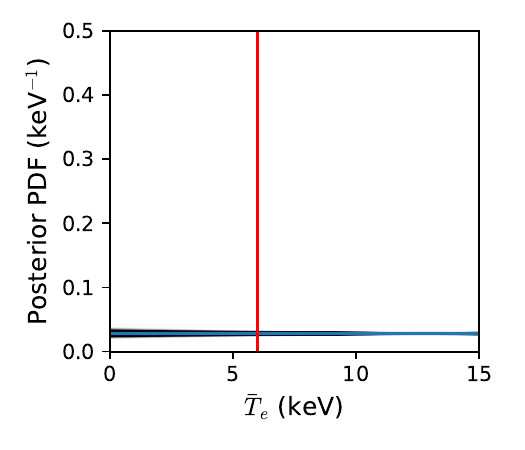} }
  \end{subfloat}
    \caption{The posteriors for 1000 mock analyses of a very faint cluster sample  data vector. The lines correspond to the same quantities as \cref{fig:example_largeAmp}. The signal in this sample is significantly below the noise, thus the data provides no information on the temperature. With the flat prior (left) the posterior peaks at zero, the temperature with the lowest measurement uncertainty. With the uninformative prior (right) we get the expected behavior - when the data have no information on the temperature, the temperature is unconstrained. }
    \label{fig:example_smallAmp} 
\end{figure*}

The results are shown in \cref{fig:example_largeAmp} and \cref{fig:example_smallAmp}. We see that in both cases the uniform temperature prior prefers lower temperatures than the true temperature, driven by the fact that, for our instrumental configuration, lower temperatures are more tightly constrained. This is the case even for the second case, with a very small signal. In this case the data contain no information about the temperature as the measurements are completely noise dominated. The uninformative prior instead has a uniform posterior. 

In summary, we choose the uninformative prior as it more accurately recovers the input temperature when the data are constraining and returns a uniform posterior when the data are not constraining. 

\bibliographystyle{prsty.bst}
\bibliography{rSZ,Planck_bib}
\newpage
\end{document}